\documentclass[twocolumn,onecolappendix]{aastex631}

\usepackage{bm,amsmath,esint}

\shorttitle{High energy pulsars}
\shortauthors{Hakobyan et al.}
\graphicspath{{./}{figures/}}

\newcommand{\NEW}[1]{{#1}}
\newcommand\runit{{\hat{\bm{r}}}}

\begin{document}

\title{Magnetic energy dissipation and $\gamma$-ray emission in energetic pulsars}

\author[0000-0001-8939-6862]{Hayk Hakobyan}
\affiliation{Department of Astrophysical Sciences, Princeton University, 4 Ivy Ln, Princeton, NJ 08540, USA}
\affiliation{Princeton Plasma Physics Laboratory, 100 Stellarator Rd, Princeton, NJ 08540, USA}
\affiliation{Physics Department \& Columbia Astrophysics Laboratory, Columbia University, 538 West 120th Street, New York, NY 10027, USA}

\author[0000-0001-7801-0362]{Alexander Philippov}
\affiliation{Center for Computational Astrophysics, Flatiron Institute, 162 Fifth Avenue, New York, NY 10010, USA}
\affiliation{Department of Physics, University of Maryland, College Park, MD 20742, USA}

\author[0000-0001-9179-9054]{Anatoly Spitkovsky}
\affiliation{Department of Astrophysical Sciences, Princeton University, 4 Ivy Ln, Princeton, NJ 08540, USA}

\begin{abstract}

Some of the most energetic pulsars exhibit rotation-modulated gamma-ray emission in the $0.1$ to $100$ GeV band. The luminosity of this emission is typically $0.1\text{-}10\%$ of the pulsar spin-down power (gamma-ray efficiency), implying that a significant fraction of the available electromagnetic energy is dissipated in the magnetosphere and reradiated as high-energy photons. To investigate this phenomenon we model a pulsar magnetosphere using 3D particle-in-cell simulations with strong synchrotron cooling. We particularly focus on the dynamics of the equatorial current sheet where magnetic reconnection and energy dissipation take place. Our simulations demonstrate that a fraction of the spin-down power dissipated in the magnetospheric current sheet is controlled by the rate of magnetic reconnection at microphysical plasma scales and only depends on the pulsar inclination angle. We demonstrate that the maximum energy and the distribution function of accelerated pairs is controlled by the available magnetic energy per particle near the current sheet, the magnetization parameter. The shape and the extent of the plasma distribution is imprinted in the observed synchrotron emission, in particular, in the peak and the cutoff of the observed spectrum. We study how the strength of synchrotron cooling affects the observed variety of spectral shapes. Our conclusions naturally explain why pulsars with higher spin-down power have wider spectral shapes and, as a result, lower gamma-ray efficiency.

\end{abstract}

\section{Introduction}
\label{sec:psr-intro}

In $\gamma$-ray pulsars a significant fraction of the spin-down power (between $0.1\%$ and $10\%$) is converted into high energy photons \citep{2013ApJS..208...17A}. This suggests that somewhere in the magnetosphere the Poynting flux radiated by the rotating star is efficiently dissipated and converted into {the energy of pairs populating the magnetosphere} and, ultimately, to $\gamma$-rays. {The equatorial current sheet beyond the light cylinder, where magnetic reconnection takes place, is a likely candidate where this energy extraction can take place}. The emergence of the current sheet in plasma-filled magnetospheres of neutron stars was predicted in the force-free formulation \citep{1999ApJ...511..351C, 2005PhRvL..94b1101G, 2006MNRAS.368.1055T}, while kinetic plasma simulations allowed to model them self-consistently, capturing the process of magnetic reconnection \citep{2014ApJ...785L..33P, 2014ApJ...795L..22C, 2015MNRAS.448..606C, 2015MNRAS.449.2759B, 2018ApJ...857...44K, 2020A&A...642A.204C}. 2D simulations of axisymmetric magnetospheres were able to achieve significant separation of scales between the macroscopic extent of the current layer and the microscopic plasma-kinetic scale. However, the dynamics of a 2D reconnecting current sheet may be different from that in 3D. Current layers in 3D are prone to kinetic instabilities, such as the kink instability, which may disrupt the layer, interfering with the magnetic reconnection and tearing instability, and potentially leading to the suppresion of dissipation rate \citep{2021ApJ...919..111G, 2021JPlPh..87f9013W, 2021ApJ...922..261Z}. Neutron stars which have magnetic axes misaligned with respect to their rotation axes have intrinsically non-axisymmetric magnetospheres and, as a result, more complex structures of current layers {which can only be studied in 3D}. 


In global 3D simulations large scale separation is numerically challenging and expensive, and yet critically necessary to resolve the hierarchical chain of plasmoids that occur in the magnetic reconnection and mediate the particle acceleration process. As a result, while previous 3D simulations properly capture the general structure of the magnetosphere, the complex dynamics of the current layer is still largely unexplored due to the rather limited scale separation in prior works. In this work we make an attempt to properly capture the microphysics of the current sheet in 3D PIC simulations by extending the separation of macro-to-micro scales to $\gtrsim 100$. We achieve this by employing a hybrid approach for particle pushing in our simulations, where the motion of particles in highly magnetized regions is reduced to that of their guiding centers, while in the accelerating regions the full motion is recovered \citep{2020ApJS..251...10B}. This approach also enables us to include a strong self-consistent synchrotron radiation-reaction force acting on the subsect of particles for which gyration is resolved. 



{The goal of the present work is to quantify the amount of magnetic energy dissipation, determine what plasma parameters control the average and the maximum energy that particles gain during the acceleration process, and what role does the strong synchrotron cooling, present in most of the energetic pulsars, play.} We begin in section~\ref{sec:psr-magnetosphere} with a brief introduction to the 3D structure of the magnetosphere, and the numerical setup we employ. We then describe the energy dissipation process in the reconnecting current layer and demonstrate that the rate of this process controls the overall dissipation rate in the magnetosphere (section~\ref{sec:psr-dissipation}). In section~\ref{sec:psr-radiation} we study the particle acceleration and high-energy radiation {in the regimes of dynamically strong and weak synchrotron cooling.} Section~\ref{sec:psr-observ} summarizes our main findings, putting them in the context of {\it Fermi} observations. {In particular, we show how the interplay of radiative losses and particle acceleration in magnetic reconnection explains the observed diversity of $\gamma$-ray spectra. }

\section{Plasma-filled pulsar magnetospheres}

\label{sec:psr-magnetosphere}
Pair cascade near the neutron star surface has long been thought to populate the neutron star magnetospheres with pair plasma \citep{1971ApJ...164..529S,1975ApJ...196...51R}. {When the magnetosphere has enough plasma supply, $n_e \gtrsim \rho_{\rm GJ}/|e|$, its structure relaxes to the force-free (FF) solution, where the electric field component parallel to the magnetic field vanishes everywhere, $\bm{E}\cdot\bm{B}=0$. Charge density necessary to sustain FF solution is called the \emph{Goldreich-Julian} (GJ) density, $\rho_{\rm GJ} = \bm{\Omega} \cdot\bm{B}/(2\pi c)$ \citep{1969ApJ...157..869G}. For the most energetic pulsars pair-producing cascade is highly efficient {and can populate the magnetosphere with pair multiplicities up to $n_e |e|/\rho_{\rm GJ}\sim 10^4$} \citep{2015ApJ...810..144T, 2019ApJ...871...12T}. In this work we only consider neutron star magnetospheres with the abundant pair plasma supply, where deviations from the FF solution are at the scale of the local plasma skin-depth, $d_{e}$.}

\subsection{Numerical setup}

We employ radiative particle-in-cell (PIC) algorithm implemented in the \texttt{Tristan-MP v2} code designed by \cite{tristanv2} to simulate the 3D dynamics of the entire magnetosphere. Synchrotron radiation drag force is included in the equations of motion for particles in the Landau-Lifshitz form. The magnitude of the force is artificially enhanced with respect to the Lorentz force, so the most energetic particles lose a substantial amount of energy on the gyration time scales. Similar approach for modeling synchrotron drag has been used in the previous works \citep[e.g.,][]{2016MNRAS.457.2401C}.

Our simulations start with an empty Cartesian domain of the size $\sim (5 R_{\rm LC})^3$, where $R_{\rm LC}=c/\Omega$ is the light cylinder of the magnetosphere, and $\Omega$ is the spin frequency of the neutron star. The star itself is modeled as a perfectly conducting rotating sphere in the center of the domain \citep[similar to][]{2015ApJ...801L..19P}. Dipolar magnetic field is imposed as a boundary condition near the surface of the sphere, as well as the initial condition in the whole domain. The angle between the magnetic axis and the rotation axis is further denoted by $\chi$. Near outer boundaries all the electromagnetic field components are damped to zero, and the particles are allowed to leave. Particle distribution is sampled on average by $\sim 10$ macroparticles per grid cell (PPC), with fewer PPC further from the star.\footnote{{This number varies from around $\sim 1000$ in the small region near the surface of the star to a few-to-ten in the bulk of the magnetosphere, and $\sim 100$ in the current layer.}} To mitigate the numerical artifacts from finite PPC we employ digital filtering of the deposited currents with a stencil of size $2$. We also tested the setup with smaller resolution simulation and $10$ times more PPC, arriving to the same results.


To fill the magnetosphere with plasma we mimic the pair production process near the surface of the star by injecting pair plasma in a small spherical shell of size $\Delta r$ at a rate proportional to the local GJ density: $\Delta n(\theta,\phi)/\Delta t = f_{\rm inj}|\rho_{\rm GJ}(\theta,\phi) / e|(c / \Delta r)$. Dimensionless parameter $f_{\rm inj}$ controls the injection multiplicity. The exact value of this number is unimportant, as long as enough plasma is injected to establish the force-free solution (we typically set $f_{\rm inj}=1$). We also give a marginal initial velocity to the newly injected particles along the local magnetic field lines  (typically a Lorentz factor of $\gamma\approx 2$).\footnote{It is important to mention that our technique of providing the magnetosphere with fresh plasma is different from the more self-consistent methods used in previous works \citep{2014ApJ...795L..22C, 2015ApJ...801L..19P}. To simplify the simulation as well as to reduce the number of input parameters, we do not model the full pair production process. However, as was demonstrated in the earlier works \citep{2015MNRAS.448..606C, 2015MNRAS.449.2759B}, the structure of the outer magnetosphere is unaffected by the pair injection prescription near the polar cap.}

To be able to resolve the plasma skin depth, $d_e$, everywhere except for the surface of the star we greatly reduce the scale separation compared to realistic pulsars. The radius of the star is resolved by $75\Delta x$ (we will denote $\Delta x$ as the size of the grid cell), while the size of the light cylinder is  $R_{\rm LC}\sim 440\Delta x$ (or $\sim 6$ times the radius of the star). The strength of the magnetic field at the surface, $B_*$, which also rescales the Goldreich-Julian density, is chosen in such a way that $d_e^{\rm LC}\sim \text{few}~\Delta x$, where $d_e^{\rm LC}$ is the plasma skin depth near the light cylinder. The scale separation between macroscopic and microscopic (plasma kinetic) length scales is, thus, at most $\sim200$ in our highest resolution simulation (cf. almost $8$ orders of magnitude scale separation in realistic pulsars). Large separation of scales ensures that the growth of microscopic plasma instabilities that develop at kinetic timescales (inverse plasma frequency, $\omega_{{\rm p}e}^{-1}$) is much faster than the dynamical timescale characterized by the rotation period, $P$. Localized 2D simulations \citep[see, e.g.,][]{2016ApJ...816L...8W} show that at scale separation $\gtrsim 100$ kinetic instabilities establish an asymptotic regime, which justifies our choice of $R_{\rm LC} / d_e^{\rm LC} \sim \omega_{{\rm p}e}^{\rm LC}/\Omega \sim 200$. The choice of these parameters yields the total size of the simulation domain $(2200\Delta x)^3$.



In regions where the magnetic field is strong (close to the surface of the star) we significantly underresolve particle gyration timescale: $\omega_B\Delta t/\gamma\gg 1$, where $\omega_B = |e|B/m_e c$, and $\gamma\approx$ few). To avoid associated numerical errors we employ a coupled guiding-center/Boris algorithm for solving particle equations of motion \citep{2020ApJS..251...10B}. This approach allows to ignore gyrations of most of the particles in the bulk of the magnetosphere where the gyration is underresolved, reducing their motion to that of their guiding centers. At the same time, switching to conventional Boris pusher in regions where $E/B > 0.95$ allows to recover the full equation of motion for high-energy particles in regions with vanishing magnetic field (i.e., current sheets).\footnote{More details on how the coupled particle pusher algorithm works can be found in the appendix~\ref{sec:psr-appendixB}.} By ignoring the synchrotron drag force for particles in the guiding center regime, we can also avoid having numerical errors in the strong cooling regime, when the synchrotron cooling algorithm heavily relies on the resolution of gyration orbit.

Overall, the three dimensionless parameters that we fix independently in our simulations are:
\begin{itemize}
  \item $R_{\rm LC} / d_e^{\rm LC}\sim 100\text{-}200$: the ratio of the size of the light cylinder and the plasma skin depth at a corresponding GJ plasma density, we refer to this as the \emph{scale separation} of our simulation;
  \item $R_* / \Delta x\approx 75$: the number of simulation cells per stellar radius, which we call the \emph{resolution} of our simulation;
  \item $R_{\rm LC} / R_*\approx 6$: the size of the light cylinder w.r.t. the size of the star.
\end{itemize}
\noindent In addition we also have the obliquity angle, $\chi$, which we vary from $0^\circ$ to $90^\circ$, and the cooling strength, which we describe in details further.

\subsection{Steady-state numerical solution}

\begin{figure*}[htb]
\centering
\includegraphics[width=2\columnwidth]{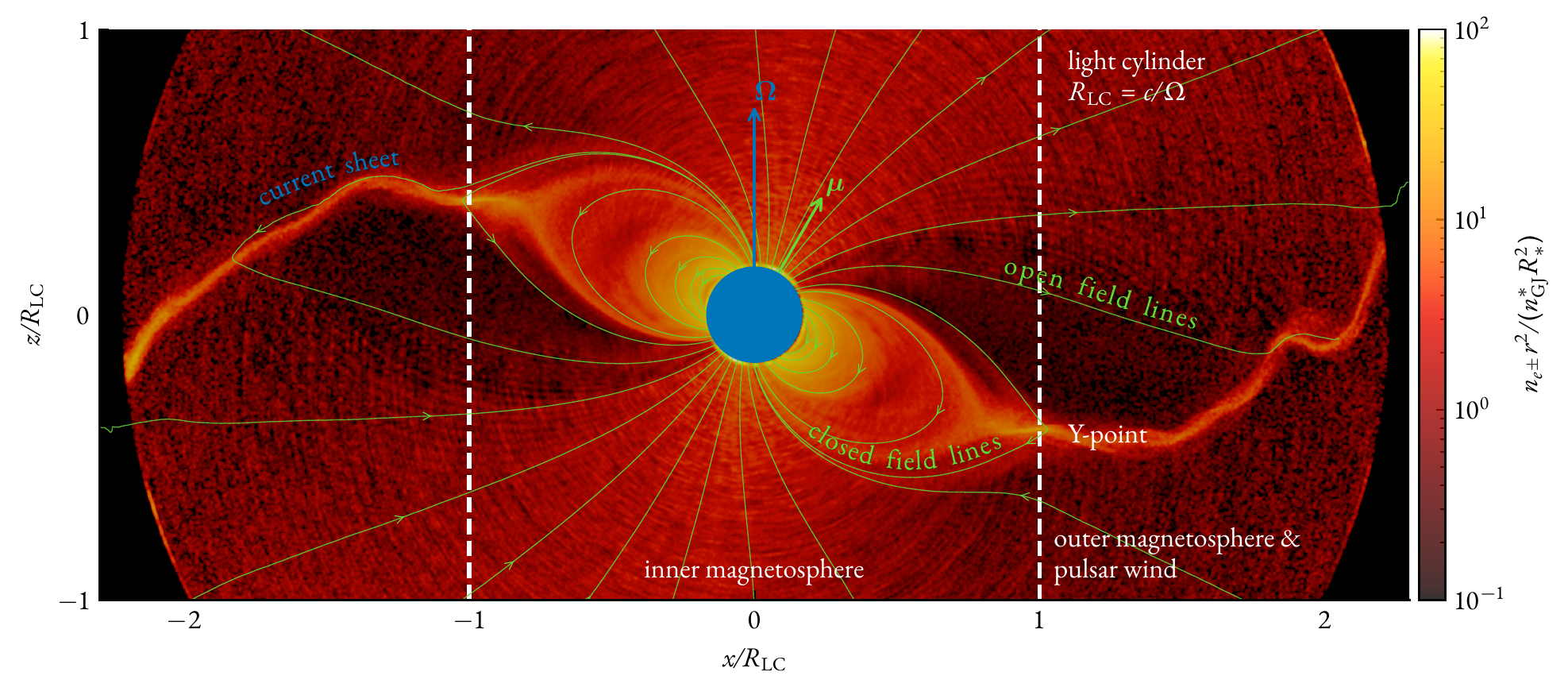}
\caption{Poloidal slice of the 3D plasma density {normalized to $n_{\rm GJ}^* R_*^2 / r^2$} from a simulation of an inclined rotator with $\chi=30^\circ$ {(here $n_{\rm GJ}^*$ is the GJ density near the pole of the star)}. The surface of the neutron star and its rotation axis are shown in blue. Magnetic field lines traced from the surface of the star, as well as the direction of the magnetic moment, are shown in green. The light cylinder, $R_{\rm LC}$, shown with white dashed lines, separates the inner magnetosphere from the outer magnetosphere and the wind. Current sheet originating near the Y-point and spreading into the outer magnetosphere is clearly visible. }
\label{fig:psr-pulsardraft}
\end{figure*}

In all of our simulations, after a brief transient lasting for about 1 rotation period of the star, $P$, a steady-state structure is established which closely resembles the FF solution. The snapshot of the steady-state for the neutron star with an inclination angle $\chi=30^\circ$ is outlined show in Figure~\ref{fig:psr-pulsardraft}. Plasma close to the surface corotates with the neutron star, flowing along almost poloidal magnetic field lines that start and end at the surface of the star. This corotation extends up to $R_{\rm LC}=c/\Omega$ from the rotation axis ($\Omega$ being the rotation frequency of the neutron star), the \emph{light cylinder}. Outside the light cylinder plasma can no longer corotate with the star, and pairs slide along spiral magnetic field lines beyond the $R_{\rm LC}$, moving almost radially outward. While in the inner magnetosphere (inside the light cylinder) the magnetic field is almost purely poloidal, in the outer magnetosphere it has a toroidal components. 

The rotation of the magnetosphere imposes a poloidal electric field {in the wind zone}, $E_{\theta} \sim B_{\phi}$. As a result, pulsar radiates electromagnetic energy in the form of a radial Poynting flux: $(4\pi / c)\bm{S} = \bm{E}\times\bm{B} = E_{\theta} B_{\phi} \runit$. This flux, integrated over a sphere that encloses the light cylinder, is the \emph{spin-down energy} of the neutron star, often denoted as $\dot{E}$, which for an aligned rotator reads:

\begin{equation}
\label{eq:psr-edot}
\begin{split}
    \dot{E} \equiv L_0 &= \oiint\limits_{r=R_{\rm LC}} \bm{S} \cdot d\bm{a} \\
    &= 2\pi \frac{B_*^2 R_*^3}{P}\left(\frac{R_*}{R_{\rm LC}}\right)^3,
\end{split}
\end{equation}
where $B_*$ is the magnetic field strength near the stellar surface at the equator. 

\begin{figure*}[htb]
  \centering
  \includegraphics[width=2\columnwidth]{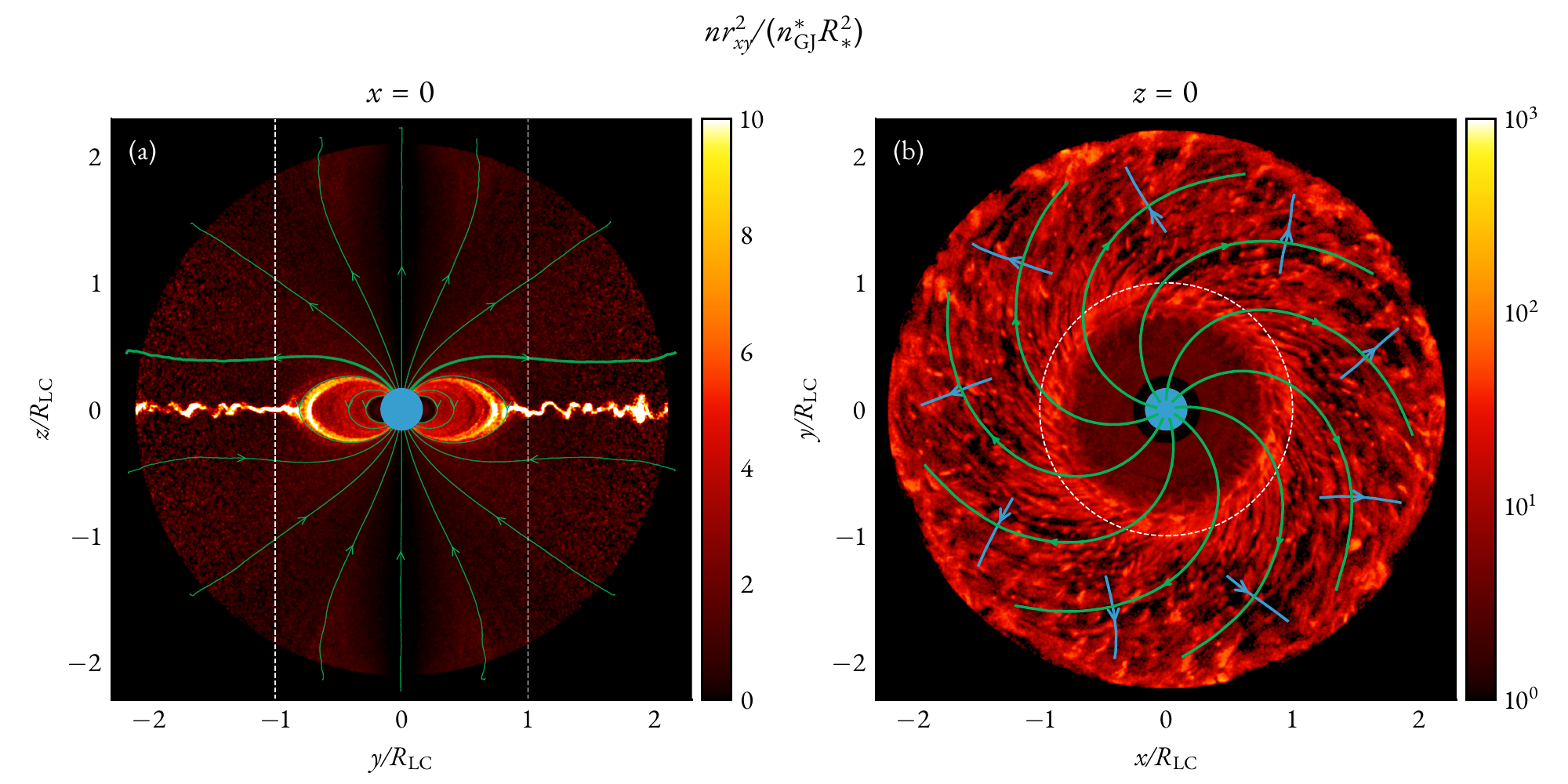}
  \caption{(a): poloidal slice of the plasma density from a simulation of an aligned rotator (\texttt{R75\_ang0}, after $2$ full rotations of the neutron star). Magnetic field lines originating at the stellar surface are shown with green lines. White dashed lines indicate the light cylinder. (b): equatorial slice of the plasma density. As in (a), green lines show the direction of the magnetic field. White dashed circle corresponds to the light cylinder. Blue arrows trace the direction of the equatorial current in the equatorial sheet. Ripples in density are caused by the combined effect of two instabilities (tearing and kink) in the current sheet. In the slice (a) the drift-kink instability of the current sheet is apparent. Slight collimation of field lines near the boundaries close to the poles in (a) is caused by imperfect absorbing boundary conditions in the aligned case. This feature, not present in simulations of misaligned pulsars, however, does not affect the global solution, or the dissipation in the current sheet.}
  \label{fig:psr-pulsarslice}
 \end{figure*}


Opposing field lines from the northern and southern hemispheres of the neutron star are divided in the outer magnetosphere by a \emph{current sheet}. The structure of the equatorial current sheet is better visible in the simulation with no obliquity shown in Figure~\ref{fig:psr-pulsarslice}a,b (hereafter we refer to this simulation as \texttt{R75\_ang0}, since $\chi=0^\circ$, and $R_*=75\Delta x$).\footnote{Simulations with the same basic parameters and other obliquity angles are, correspondingly, \texttt{R75\_ang20}, \texttt{R75\_ang60}, and \texttt{R75\_ang90}.} Cartesian coordinates are normalized to the $R_{\rm LC}$, and $(0, 0, 0)$ is the middle of the box, where the center of the neutron star is. Figure~\ref{fig:psr-pulsarslice}a shows the slice of the simulation in the $x=0$ plane, while Figure~\ref{fig:psr-pulsarslice}b shows the slice in the $z=0$ plane (in this simulation $\chi=0^\circ$). Color indicates the plasma number density compensated by the cylindrical radius squared, $n r_{xy}^2$; its value is normalized by the corresponding GJ density at the surface of the star times the radius of the star squared, $n_{\rm GJ}^* R_*^2$. 


From Figure~\ref{fig:psr-pulsarslice} it is evident that the total plasma density close to the current sheet is few to ten times larger than the local GJ density, which means the magnetosphere has enough plasma to screen the accelerating electric field almost everywhere. If the parallel electric field were screened everywhere, magnetic energy dissipation in the magnetosphere would not be possible, and the integral in Eq.~\eqref{eq:psr-edot} would be constant with $r$, $L(r)\equiv \oiint_r \bm{S} \cdot d\bm{a}=const$, {yielding no $\gamma$-ray emission}. 

    
The accelerating electric field can survive in microscopic sub-regions of the equatorial current sheet. The characteristic scale of these regions does not exceed few-to-tens of plasma skin depths, $d_e= c/\omega_{{\rm p}e}$, where $\omega_{{\rm p}e}$ is the plasma frequency for relativistic electron-positron pairs in the current sheet. For the Crab pulsar this parameter close to the light cylinder is of the order of a few centimeters. The light cylinder, on the other hand, is almost $8$ orders of magnitude larger (for the Crab it is $\sim 1500$ kilometers). {To properly model the dynamics of these sub-regions first-principles PIC simulations are necessary.}


\section{Dissipation of spin-down energy}
\label{sec:psr-dissipation}

The structure of the magnetosphere in PIC, as was demonstrated in the previous section, is very similar to that in force-free. This should come as no surprise, since in our simulations $\bm{E}_\parallel$ is perfectly screened with abundant plasma, and magnetized particles follow field lines everywhere {except for the equatorial current sheet where the magnetic field vanishes.} The key difference from FF solutions is the dynamics of the current sheet. In ideal FF the energy dissipation in the current sheet is due to the finite resolution of the grid, and is thus numerical in its nature. PIC algorithm, on the other hand, enables us to ``resolve'' this dissipation on plasma kinetic scales, by capturing the reconnection of magnetic field lines from first principles. 


\subsection{Kinetic instabilities in the current sheet}

The equatorial current sheet highlighted in Figure~\ref{fig:psr-pulsarslice} is not laminar even in the aligned case, $\chi=0^\circ$. Rather, it is prone to kinetic instabilities that develop on microscopic timescales comparable to $\omega_{{\rm p}e}^{-1}\ll \Omega^{-1}$. \emph{Drift-kink instability} displaces the current sheet in $z$-direction, which results in the undulation of the sheet at a certain saturated amplitude (shown in the poloidal slice of Figure~\ref{fig:psr-pulsarslice}b; also see \citealt{2014ApJ...785L..33P, 2015MNRAS.448..606C}). The wave-vector of this perturbation is perpendicular to the local magnetic field and is parallel to the local current. 


More importantly, the current sheet also experiences \emph{tearing instability} due to relativistic reconnection of magnetic field lines from the upstream \citep[see, e.g.,][]{2017A&A...607A.134C, PSAS18, 2020A&A...642A.204C}. As a result of this process, magnetic islands (\emph{``plasmoids''}) are formed containing hot plasma energized in the reconnection process. In-between the plasmoids there are magnetic nulls -- regions where the magnetic field lines tear, and energy dissipation happens, the \emph{``x-points.''}

\begin{figure*}[htb]
\centering
\includegraphics[width=2\columnwidth,trim={10 10 10 5},clip]{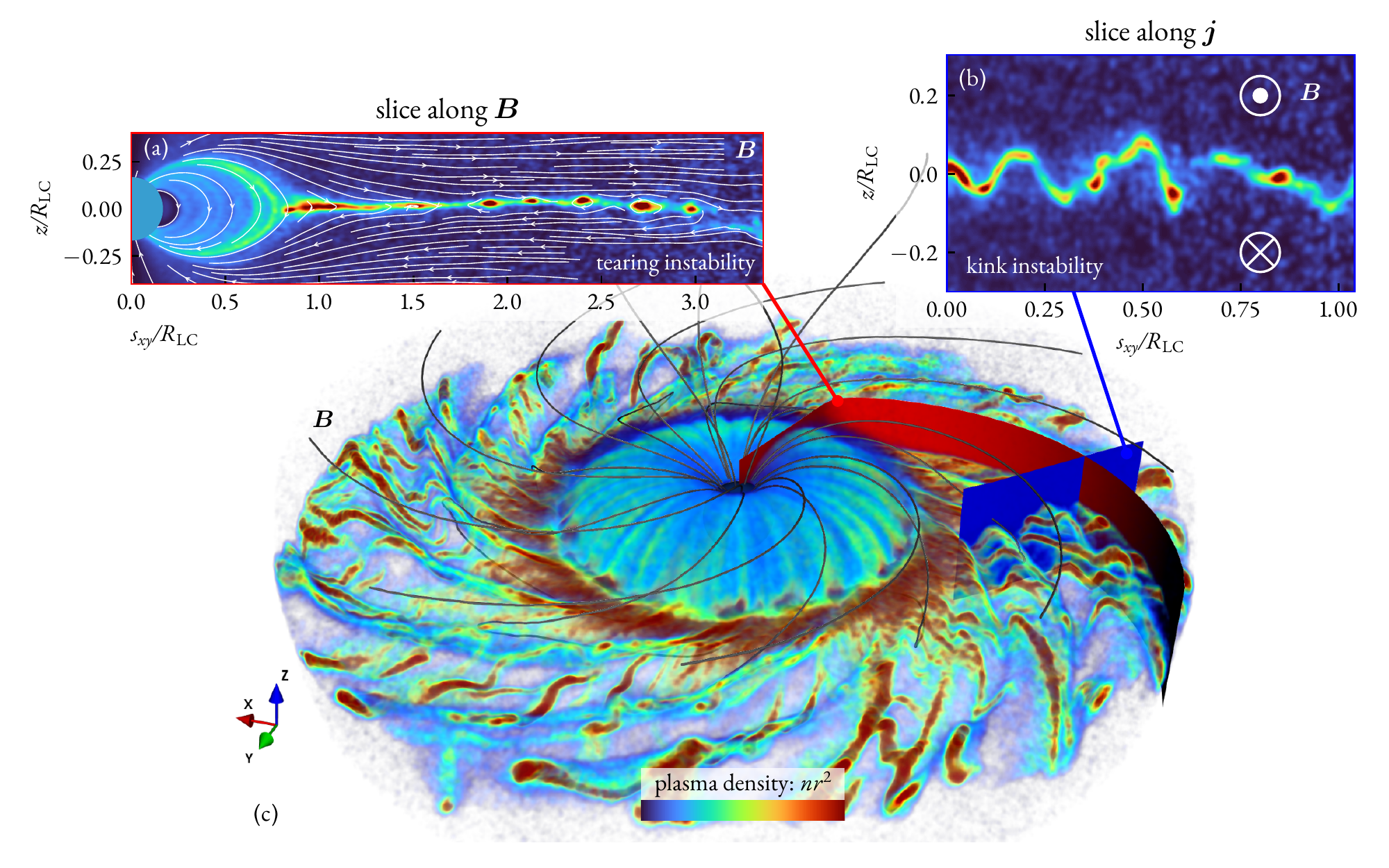}
\caption{3D rendering of the plasma density (compensated by $r^2$) from a simulation of an aligned rotator (\texttt{R75\_ang0}). 3D rendering (panel c) is accompanied by two slices (a and b, also visible in 3D as red and blue surfaces). One of the slices (a) is orthogonal to the equator and curves along the upstream magnetic field lines, while the second one (b) is along the direction of the current perpendicular to the magnetic field. Overdense elongated tubes in panel (c) are the 3D flux ropes, the plasmoids, which are produced as a result of the non-linear tearing instability. In a 2D slice (a) the dynamics of both the current sheet and the plasmoids look very similar to those in isolated current sheet simulations. In panel (b) the drift-kink instability is visible, which deforms the current sheet in the direction perpendicular to the direction of the tearing instability. An animated version of this figure is available at the following link: \url{https://youtu.be/-YXJ4yTlhWw}.}
\label{fig:psr-pulsar3d}
\end{figure*}

Figure~\ref{fig:psr-pulsar3d}c shows a 3D rendering of the plasma density as well as two slices of the same quantity: one along the magnetic field lines (Figure~\ref{fig:psr-pulsar3d}a indicated with a red surface in panel \ref{fig:psr-pulsar3d}c), and the other one -- perpendicular to field lines, along the direction of the current (Figure~\ref{fig:psr-pulsar3d}b indicated with a blue surface). In Figure~\ref{fig:psr-pulsar3d}b one can see the undulation of the current sheet due to drift-kink instability. The reconnection of magnetic field lines and tearing instability, on the other hand, are better visible in Figure~\ref{fig:psr-pulsar3d}a; overdense regions in the sheet are the plasmoids, which in 3D look like tubes elongated almost radially (Figure~\ref{fig:psr-pulsar3d}c).

\begin{figure}[htb]
\centering
\includegraphics[width=\columnwidth,trim={10 10 10 5},clip]{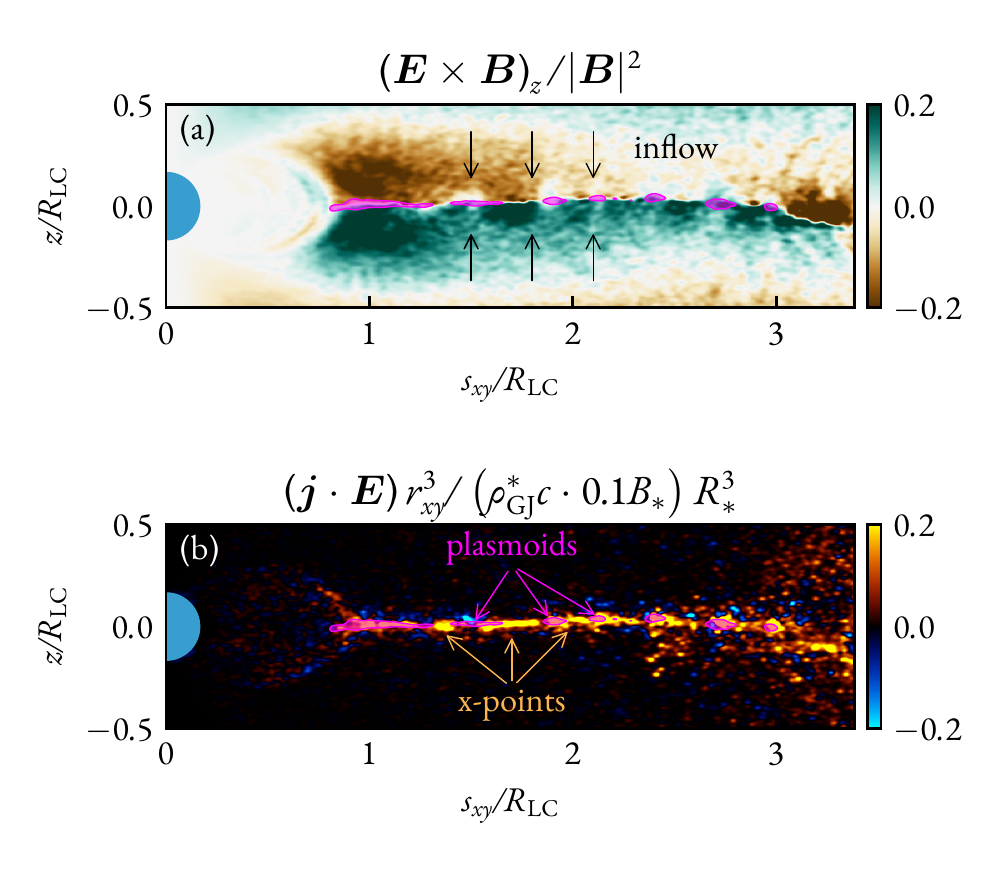}
\caption{Two panels show different quantities in the same slice as Figure~\ref{fig:psr-pulsar3d}a. (a) shows the  $\bm{E}\times\bm{B}$ inflow rate into the current sheet; this corresponds to the rate at which the reconnection of magnetic field lines occur, roughly, $\beta_{\rm rec}\sim 0.1$. In panel (b) we plot the work done by the electric field (compensated by $r_{xy}^3$), which also indicates the electromagnetic energy dissipation rate. The latter is localized in the thin equatorial current sheet. On both panels we overplot the plasmoids in magenta (their boundaries are selected as contours of large overdensities clearly visible in Figure~\ref{fig:psr-pulsar3d}a).}
\label{fig:psr-reconnection}
\end{figure}

The dynamics of the reconnecting current sheet in slice \ref{fig:psr-pulsar3d}a is very similar to a 2D Harris sheet, except for the fact that the upstream plasma moves {along the magnetic field lines} with bulk Lorentz factor $\Gamma\sim \mathcal{O}(1)$ (in reality this is expected to be $\Gamma\sim \mathcal{O}(100)$; see, e.g., discussion by \citealt{2020A&A...642A.204C}).\footnote{In addition to the motion along the field lines particles also experience an $\bm{E}\times\bm{B}$ drift. However, the Lorentz factor associated with the drift is rather small within a few light cylinders from the star.} In Figure~\ref{fig:psr-reconnection}a,b we take the slice along the upstream magnetic field lines (similar to \ref{fig:psr-pulsar3d}a) where we plot quantities crucial for understanding the energy dissipation in the current sheet: the vertical component of the $\bm{E}\times\bm{B}$ drift velocity in units of $c$, and the energy dissipation rate, $\bm{j}\cdot\bm{E}$. In the upstream (above and below the current sheet) the plasma motion is strictly constrained by magnetic field lines, which coincide with the plane of the Figure~\ref{fig:psr-reconnection}a,b. Cold plasma and opposing magnetic field lines are brought together by an $\left(\bm{E}\times\bm{B}\right)$ drift (shown in Figure~\ref{fig:psr-reconnection}a); the current sheet then becomes unstable and tears producing plasmoids (shown with magenta contours). The dimensionless rate of inflow is set by the reconnection; this value can be directly measured from the simulations. For our simulations this value is close to
\begin{equation}
    \beta_{\rm rec} \equiv \frac{\left(\bm{E}\times\bm{B}\right)_{\rm in}}{|\bm{B}|^2} \approx 0.1\text{-}0.2,
\end{equation}
where the subscript ``in'' corresponds to the velocity component directed into the current sheet. This value is consistent with isolated 2D/3D simulations of the Harris sheet, and has been demonstrated to be independent of any numerical or physical parameters (such as the number of macroparticles per cell, the resolution, or the extent of the box) \citep[see, e.g.,][]{2018MNRAS.473.4840W}.

\subsection{Magnetic energy dissipation via reconnection}

\begin{figure}[htb]
\centering
\includegraphics[width=\columnwidth,trim={10 10 10 5},clip]{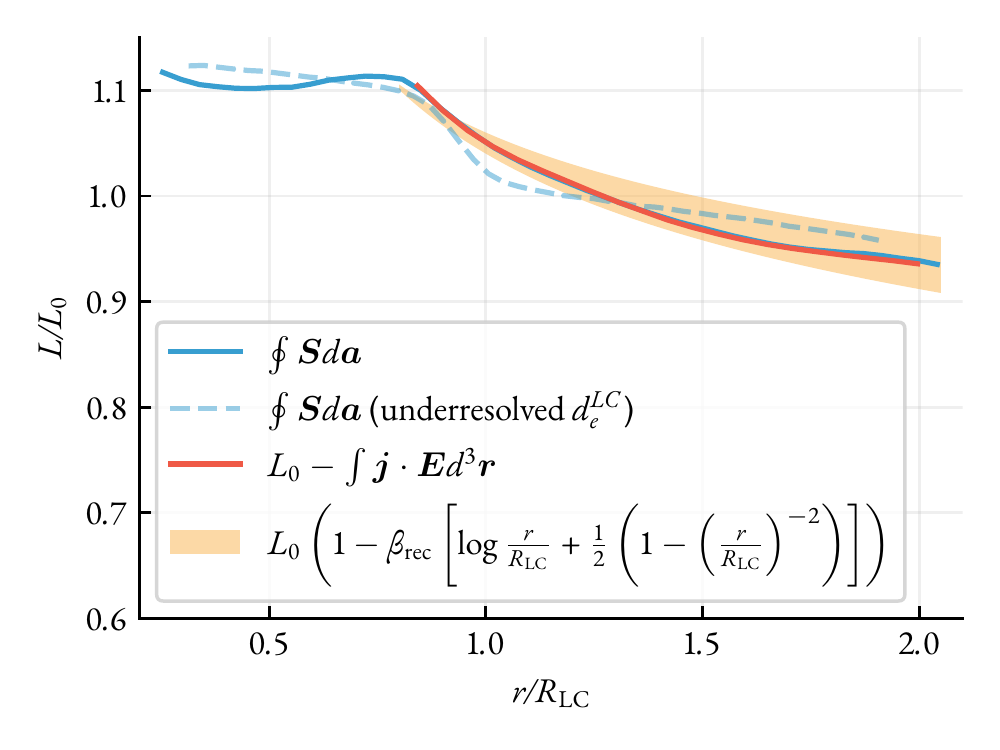}
\caption{Dissipation of the Poynting flux in the outer magnetosphere from a simulation of an aligned rotator. The blue line is the direct measurement of the Poynting flux vs $r$ for two different simulations: {\texttt{R75\_ang0} (solid line), where $d_e^{\rm LC}\approx 3\text{-}5\Delta x$, and \texttt{R60\_ang0} (dashed line), where $d_e^{\rm LC}\approx 0.2\text{-}0.5\Delta x$.} The red line shows the $\bm{j}\cdot\bm{E}$, which accounts for the dissipation rate of the electromagnetic energy. Orange band is the theoretical fit, Eq.~\eqref{eq:psr-luminosity}, with $\beta_{\rm rec}\approx 0.09\textrm{-}0.12$. $L_0$ is computed from Eq.~\eqref{eq:psr-edot}. A slight increase of $L$ with respect to $L_0$ in the inner magnetosphere is due to the fact that the Y-point in our simulation is slightly inside the $r/R_{\rm LC} = 1$; this leads to slightly more open field lines and thus more Poynting flux. Weak sycnhrotron cooling of the particles is present in both of these simulations; however, as we show further, the effect of cooling on the reconnection rate in this high magnetization regime is negligible.}
\label{fig:psr-dissipation}
\end{figure}

Reconnecting magnetic field in the current sheet generates a non-ideal electric field in the direction of $\nabla\times \bm{B}$, which is the direction of the current in Figure~\ref{fig:psr-pulsarslice}a (in Figure~\ref{fig:psr-pulsar3d}a and Figure~\ref{fig:psr-reconnection}a,b this corresponds to the out-of-plane direction). The magnitude of the non-ideal electric field is of the order of $E_{\rm rec}\sim \beta_{\rm rec}B_{\rm up}$, where $B_{\rm up}$ is the magnetic field strength in the upstream. Because of the emergence of $\bm{j}\cdot \bm{E}$ (shown in Figure~\ref{fig:psr-reconnection}b) some of the Poynting flux radiated by the star (Eq.~\ref{eq:psr-edot}) dissipates in the magnetosphere. As seen in Figure~\ref{fig:psr-reconnection}b dissipation {is concentrated} exclusively in the thin current sheet outside the light cylinder, where the energy of the electromagnetic field is converted into the kinetic energy of plasma through magnetic reconnection. According to Poynting's theorem:

\begin{equation}
\label{eq:psr-poynting-th}
    L(r) - L_0 \equiv \oiint\limits_{R_{\rm LC}}^r \bm{S} d\bm{a} = -\iiint \bm{j}\cdot \bm{E}d^3\bm{r}.
\end{equation}
\noindent {This means that $L(r)$ is constant within the light cylinder, $r<R_{\rm LC}$ (where $\bm{j}\cdot\bm{E}=0$), but decays with radius for $r>R_{\rm LC}$.} Figure~\ref{fig:psr-dissipation} shows $L(r)$ from one of our simulations with $\chi=0^\circ$ computed using the flux of the Poynting vector (blue line), and the volume integral of $\bm{j}\cdot\bm{E}$ (red line). Orange band corresponds to an analytical model described further in this section.

To understand the slope of $L(r)$ beyond the light cylinder it is useful to build a simple analytical model of the Poynting flux dissipation in the current sheet. For $\chi=90^\circ$ rotator this model has been studied by \cite{2020A&A...642A.204C}; here we focus on $\chi=0^\circ$. As mentioned earlier, the dissipation of Poynting flux in our simulations is caused by the work done by the reconnection electric field, $\bm{j}\cdot\bm{E}$, as described by Eq.~\eqref{eq:psr-poynting-th}. The strength of the electric field generated due to magnetic reconnection at a given distance $r$ from the star is $E_{\rm rec}(r)\sim\beta_{\rm rec} B_{\rm up}(r)$. Here the reconnecting magnetic field is a combination of poloidal and toroidal components {above and below the current layer}, $B_{\rm up}(r) =\sqrt{B_{r}^2 + B_\phi^2}$, \NEW{the value of which beyond the light cylinder can be approximated as $B_{\rm up}(r)=B_{\rm LC}(R_{\rm LC}/r)\sqrt{1 + (R_{\rm LC}/r)^2}$ (rotating monopole; see, e.g., \citealt{1955AnAp...18....1D}, and \citealt{1973ApJ...180L.133M}). Here and further, $B_{\rm LC}\equiv B_*(R_*/R_{\rm LC})^3$.} Since $(4\pi/c)\bm{j}=\nabla\times\bm{B}$, the current generated during reconnection can be estimated as $j\sim c B_{\rm up} / (2\pi \delta_{\rm cs})$, where $\delta_{\rm cs}$ is the characteristic width of the current sheet. Thus, the volume integral in Eq.~\eqref{eq:psr-poynting-th} reduces to:

\begin{equation}
\label{eq:psr-luminosity-int}
    L(r) - L_0 = -2\pi\int\limits_{R_{\rm LC}}^r dr~r^2\int\limits_{\pi/2-\Delta\theta/2}^{\pi/2+\Delta\theta/2} d\theta~\sin{\theta} \left(\frac{\beta_{\rm rec} c}{2\pi \delta_{\rm cs}}B_{\rm up}^2\right).
\end{equation}
\noindent Integral over the polar angle, $\theta$, at each $r$ is accumulated in a small region near the equator ($\theta=\pi/2$) of angular size $\Delta \theta\sim \delta_{\rm cs}/r$ (which is the current sheet highlighted in Figure~\ref{fig:psr-reconnection}b).\footnote{\NEW{Notice, that the exact value of $\delta_{\rm cs}$ is irrelevant, since $\int d\theta\sin{\theta} (r / \delta_{\rm cs})\approx 2$ (for $\theta\in(\pi/2-\delta_{\rm cs}/r,\pi/2+\delta_{\rm cs}/r)$), as long as $\delta_{\rm cs}\ll r$.}} Integrating Eq.~\eqref{eq:psr-luminosity-int} then yields

\begin{equation}
\label{eq:psr-luminosity}
    \frac{L(r)}{L_0} = 1 - \beta_{\rm rec}\left(\ln{\frac{r}{R_{\rm LC}}}
    +\frac{1}{2}\left(1 -\left[\frac{r}{R_{\rm LC}}\right]^{-2}\right)\right),
\end{equation}

\noindent \NEW{where we also used $L_0\equiv B_{\rm LC}^2 R_{\rm LC}^2 c$.} The logarithmic term in this expression corresponds to the dissipation of toroidal field, $B_\phi$, while the second term corresponds to the dissipation of poloidal field, $B_{r}$. At large distances, $r\gg R_{\rm LC}$ the first term prevails, as the field is almost purely toroidal. However, for the small region, $R_{\rm LC}<r<2R_{\rm LC}$, where $\gamma$-ray emission likely originates, contributions from both terms are important. In Fig.~\ref{fig:psr-dissipation} we plot the total Poynting flux, $L$, across a spherical shell of radius $r$ (blue line). Orange band in Figure~\ref{fig:psr-dissipation} corresponds to Eq.~\eqref{eq:psr-luminosity} with $\beta_{\rm rec}$ varying from $0.09$ to $0.12$ (this number can be directly measured from Figure~\ref{fig:psr-reconnection}a). From the plot it is evident that about $10\%$ of the radiated Poynting flux, $L_0$, is dissipated in the outer magnetospheric current sheet within the first $R_{\rm LC}$. The magnetic field strength weakens with distance, and particles can only radiate (via synchrotron) in GeV range within the first one-to-few $R_{\rm LC}$. Thus, the measured $\gamma$-ray luminosity, $L_\gamma$, should be directly correlated with the energy dissipated in this narrow range of radii.\footnote{In fact, in the radiative regime considered in our simulations a considerable amount (about half) of the dissipated energy is radiated away, with the other half being deposited into the acceleration of pairs.} The precise value of the dissipated energy, as we have demonstrated above, is only determined by the reconnection rate at plasma microscales.

As we demonstrate in the next section, in our simulations the plasma in the upstream of the current sheet moves along the magnetic field lines with relativistic velocities of $\Gamma\sim 10$. Naively one would expect that since the rate of reconnection is universally defined in the proper frame of the upstream, in the lab frame the measured rate should be $\Gamma$ times slower. Nevertheless, we do not observe any significant change in $\beta_{\rm rec}\sim 0.1$ (similar observations have been made by~\citealt{2020A&A...642A.204C}). To check the consistency of this result in a more controlled way we performed local 2D simulations of the current sheet, where the upstream plasma was initialized with a relativistic velocity component along the unreconnected magnetic field. From these simulations we find that the global reconnection rate in a given reference frame is controlled by the strength of the electric field in the x-points which are at rest. \NEW{Despite the motion of the upstream, reconnection rate remains constant, as long as the outflow velocities from the x-points remain relativistic. This condition is satisfied as long as the characteristic Alfv\'en four-velocity in the current sheet is much larger than the square root of the bulk four-velocity of the upstream, $\Gamma$: $u_{\mathcal{A}}\sim \sqrt{\sigma}\gg \sqrt{\Gamma}$.\footnote{More details of our findings together with some discussion can be found in the appendix~\ref{sec:psr-appendixC}.} For realistic pulsar parameters the particle four-velocity, $\Gamma$, which close to the light cylinder is mostly along the magnetic field-lines, is of the order of $10\text{-}100$, while the magnetization is close to $10^5\text{-}10^6$. This means that the reconnection is unlikely to slow down because of the bulk motion of the upstream.}


For other values of $\chi$ our results are consistent with earlier works (PIC: \citealt{2015ApJ...801L..19P}, and MHD: \citealt{2013MNRAS.435L...1T}). Namely, the spin-down luminosity near the light cylinder $\oint \bm{S}d\bm{a}\approx L_0(1 + \sin^2{\chi})$, and for larger values of $\chi$ we see inhibited dissipation in the current sheet, as the jump of the magnetic field across the current sheet is balanced primarily by the displacement current (as opposed to the conduction current) for larger inclinations~\citep[see, e.g.,][]{PSAS18}. 

Some of the previous 2D axisymmetric simulations of aligned pulsars \citep{2015MNRAS.449.2759B, 2021arXiv210903935H} reported the presence of excess dissipation near the Y-point, at the level of a few percent of $L_0$, with a lower dissipation rate at $r>R_{\rm LC}$. While the reduced dissipation rate is caused by the fact that in 2D simulations only one of the magnetic field components can be reconnected (only the poloidal one), the anomalous dissipation near the Y-point has unclear origins. In our 3D simulations we also see signs of excess dissipation in the Y-point, but only in the aligned case, and only when the skin depth near the light cylinder is marginally resolved (dashed blue line in Figure~\ref{fig:psr-dissipation}).\footnote{{We see this at resolution of $(860\Delta x)^3$, where $d_e^{\rm LC}\sim 0.5\Delta x$. For higher resolution of $(2200 x)^3$ considered, e.g., in Fig.~\ref{fig:psr-reconnection}, and $d_e^{\rm LC}\sim 5\Delta x$ we see no signs of this effect.}}

The electromagnetic energy dissipated during the reconnection in the current sheet is deposited into plasma particles. In the next section we focus on particle energization in the current sheet and its consequences for the observed high-energy emission in $\gamma$-ray pulsars.

\section{$\gamma$-ray emission}
\label{sec:psr-radiation}
Relativistic magnetic reconnection is known to produce non-thermal particle population \citep{2014ApJ...783L..21S, 2014PhRvL.113o5005G, 2016ApJ...816L...8W}; reconnection in the current sheets of neutron star magnetospheres is no exception. These energized particles are believed to radiate via synchrotron mechanism, producing the pulsed high-energy emission observed in $\gamma$-ray pulsars \citep{2016MNRAS.457.2401C,PSAS18}. {However, the synchrotron drag, the strength of which depends on the value of the magnetic field outside the current sheet, can inhibit the acceleration process, greatly affecting the outgoing photon spectrum.}


\subsection{Dynamical importance of synchrotron drag}

In the current sheets of pulsar magnetospheres the synchrotron cooling timescale for particles is always much shorter than the system-crossing timescale, characterized by the rotation period. However, it can be comparable to the acceleration time of the highest-energy particles, meaning that the relative importance of cooling can be significant at microscopic timescales. The efficiency of cooling can be conveniently quantified with a dimensionless number $\gamma_{\rm rad}$. This number corresponds to the Lorentz factor of particles for which the synchrotron drag force in the given background magnetic field $B$ is comparable to the Lorentz force from the reconnection-driven electric field $E\sim \beta_{\rm rec}B$. This condition reads: $|e|\beta_{\rm rec}B = (4/3)\sigma_T B^2 \gamma_{\rm rad}^2$, where $\sigma_T$ is the Thomson cross-section.\footnote{In terms of PIC simulations, defining dimensionless $\gamma_{\rm rad}$ is equivalent to upscaling the classical electron radius, $r_e$, to boost the relative efficiency of synchrotron losses. Also note that in this rough definition the magnetic field is considered to be exactly perpendicular to the motion of a particle. In our simulations, as well as in reality, particles flying at small pitch angles w.r.t. the magnetic field will be cooled less efficiently. Thus, $\gamma_{\rm rad}$ can be thought of as just a proxy for the average cooling efficiency for a population of isotropically distributed particles.} Particles with $\gamma \gg \gamma_{\rm rad}$ will lose their energies while being exposed to a perpendicular magnetic field component much faster than they are able to accelerate; the opposite is true for particles with $\gamma \ll \gamma_{\rm rad}$. Notice that strictly speaking this is not applicable to particles at x-points, since the magnetic field vanishes there.

\begin{figure}[htb]
\centering
\includegraphics[width=\columnwidth,trim={10 10 10 5},clip]{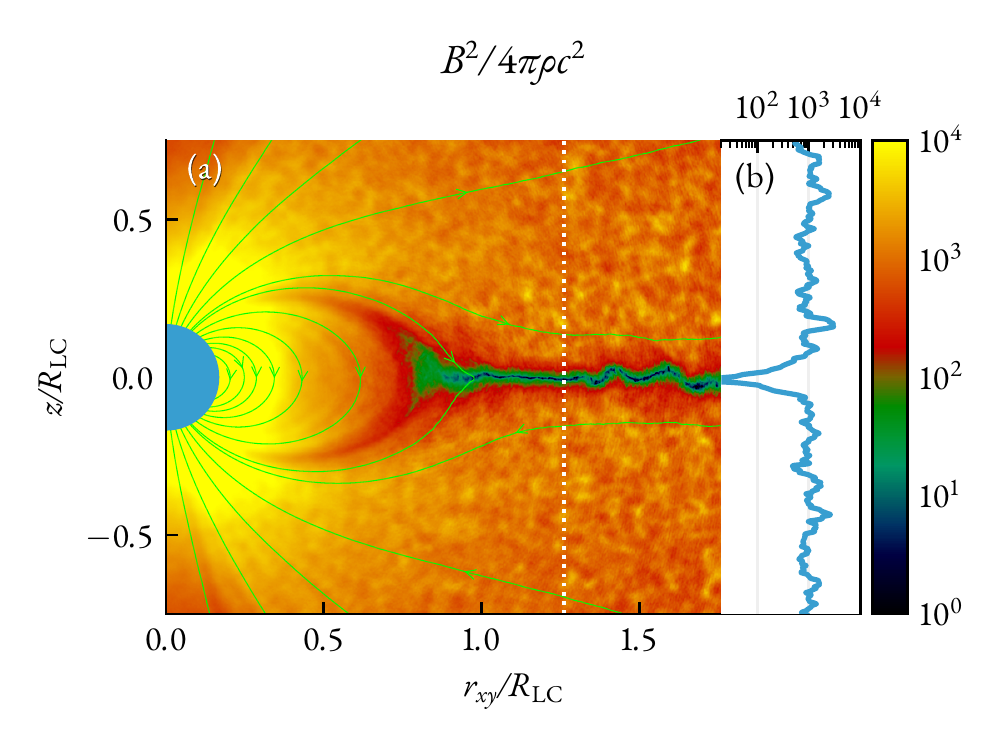}
\caption{(a): plasma magnetization parameter, $\sigma$, in a poloidal slice of the simulation \texttt{R75\_ang0\_gr2} {with weak synchrotron cooling ($\gamma_{\rm rad}^{LC}\gtrsim \sigma^{\rm LC}$; subscript \texttt{gr2} corresponds to $\gamma_{\rm rad}^{LC}/ \sigma^{\rm LC}\approx 2$).} 1D slice of the $\sigma$ across the current sheet is shown in panel (b). Magnetization of the upstream is close to $10^3$, where the background plasma density $n\sim \text{few}\cdot n_{\rm GJ}^*(R_*/r)^2$. Magnetic field lines are shown with green in panel (a).}
\label{fig:psr-sigma}
\end{figure}

The main parameter that determines the characteristic maximum energy to which particles accelerate during reconnection is the \emph{magnetization} of the upstream (unreconnected) plasma, $\sigma$. For a given background magnetic field, $B$, and number density of inflowing plasma, $n$, this dimensionless parameter is equal to twice the ratio of the magnetic energy density and the rest mass energy density of plasma:
\begin{equation}
    \sigma \equiv \frac{B^2/4\pi}{n m_e c^2}, 
\end{equation}
\noindent {with $B$ and $n$ being measured in the proper frame where plasma is at rest. In Figure~\ref{fig:psr-sigma} we show the magnetization parameter in the upstream of the reconnecting equatorial current sheet in a poloidal slice. In our typical simulations magnetization of the upstream, $\sigma^{\rm LC}$, is close to $500\text{-}1000$ as shown with a vertical slice in Figure~\ref{fig:psr-sigma}b.} This parameter, which in relativist reconnection is always $\gg 1$, determines the characteristic maximum energy to which particles can be accelerated in a single x-point (or x-line in 3D, see Figure~\ref{fig:psr-reconnection}b). If secondary reacceleration is prohibited, the non-thermal distribution function of accelerated particles extends at most to energies of a few times $\sigma m_e c^2$. Energized particles, however, may either then reenter the current sheet or become trapped in plasmoids and get accelerated again via secondary processes to even higher energies \citep{2018MNRAS.481.5687P,2021ApJ...912...48H,2021ApJ...922..261Z}. However, as we demonstrate below, in pulsar magnetospheres with realistic parameters this {process is supressed due to synchrotron losses.}

Further in this section we will refer to the case when $\gamma_{\rm rad} < \sigma$ as the \emph{strong cooling} regime, while the opposite case will be referred to as the \emph{weak cooling} regime. The value of $\gamma_{\rm rad}$ (close to the light cylinder) only depends on the magnetic field strength: $\gamma_{\rm rad}^{\rm LC}\approx 10^5\left(B_{\rm LC}/10^5~\text{G}\right)^{-1/2}$, which we know rather reliably for pulsars by extrapolating the magnetic field strength at the surface. The magnetization, $\sigma^{\rm LC}$, on the other hand, depends on plasma density near the light cylinder which is harder to constrain. However, we can estimate that approximately, by assuming that the cutoff frequency of the $\gamma$-ray photons corresponds to the synchrotron peak energy for the highest-energy particles with $E_e\sim 4 \sigma^{\rm LC} m_e c^2$: $\hbar \omega_B^{\rm LC} (4\sigma^{\rm LC})^2 \approx E_{\rm cut}$. This provides a rough empirical estimate for $\sigma^{\rm LC}$, which for the population of $\gamma$-ray pulsars varies between $10^5$ and $10^6$. Consequently, the reconnection in the current sheets of pulsars proceeds in the marginally radiative regime, where $\gamma_{\rm rad}^{\rm LC}\sim \sigma^{\rm LC}$. In the next section we quantitatively investigate how the electromagnetic energy dissipation rate, as well as the particle distribution and synchrotron spectra change depending on the ratio $\gamma_{\rm rad}^{\rm LC}/ \sigma^{\rm LC}$.


\subsection{Particle acceleration and synchrotron spectra in the radiatively cooled magnetosphere}
\label{sec:part_photon_dist}

\begin{figure*}[htb]
\centering
\includegraphics[width=2.0\columnwidth]{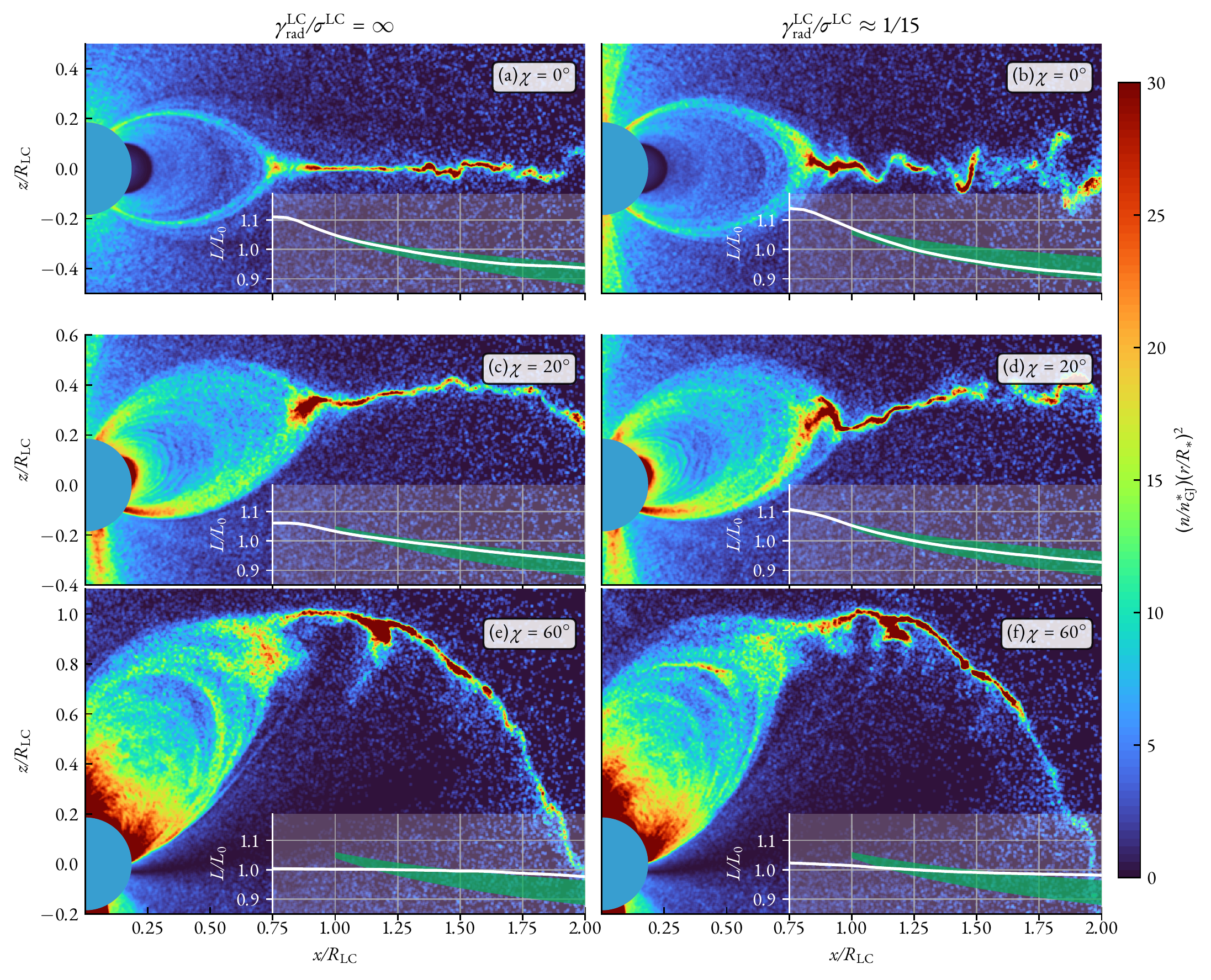}
\caption{
Density snapshots from late stages (more than 2 rotations) of six different simulations with varying obliquity angle, $\chi$, and synchrotron cooling efficiency, $\gamma_{\rm rad}^{\rm LC} / \sigma^{\rm LC}$. All of the slices are in the poloidal plane. Rows correspond to, respectively, $\chi=0^\circ$ (a, b), $20^\circ$ (c, d), and $60^\circ$ (e, f). Simulations in the left column (a, c, e) have synchrotron cooling turned off, while the ones on the right (b, d, f) have very strong synchrotron cooling, $\gamma_{\rm rad}^{\rm LC}/\sigma^{\rm LC}\approx 1/15$. Each panel also contains 1D plots of $L(r)/L_0$ (white line; {$L_0$ is the flux measured at $r=0.75 R_{\rm LC}$}) with horizontal axes, $r$, having the same scale as the $x$-axes of snapshots. Green bands correspond to the dissipation due to reconnetion estimated by the Eq.~\eqref{eq:psr-luminosity} with the rate varying between $0.8$ and $0.15$.
}
\label{fig:psr-angle_radiation}
\end{figure*}

In our simulations we keep $\sigma\gg 1$ in the outer magnetosphere (to ensure the reconnection proceeds in the ultra-relativistic regime), and vary the ratio $\gamma_{\rm rad}/\sigma$. Following the discussion earlier, this ratio determines the relative importance of the effects of synchrotron cooling on reconnection dynamics, and, as we demonstrate later, results in qualitatively different high-energy radiation spectra. {In Figure~\ref{fig:psr-angle_radiation} we show snapshots from simulations with three different inclination angles, $\chi=0^\circ,~20^\circ,~60^\circ$ (three rows), comparing two extremes of the cooling efficiency in each case: no cooling (left column, panels a, c, e), and strong cooling $\gamma_{\rm rad}^{\rm LC}/\sigma^{\rm LC} \approx 1/15$ (right column, panels b, d, f). In the same snapshot we also plot the integrated Poynting flux, $L$, as a function of radius with the $x$ axis corresponding to the radius in $R_{\rm LC}$ (shared with the $x$ axis of respective snapshots). We find that the general structure of the current sheet is largely unaffected by the strength of the synchrotron cooling. At low inclination angles $\chi\lesssim 20^\circ$ the current sheet is more intermittent with stronger cooling (compare Figure~\ref{fig:psr-angle_radiation}a and Figure~\ref{fig:psr-angle_radiation}b), but at higher inclinations the difference is negligible. The rate of magnetic reconnection and, as a result, the Poynting-flux dissipation curves, $L(r)$, are only marginally affected by the cooling strength for all values of $\chi$.}

We further closely inspect two simulations with $\chi=20^\circ$ and $\chi=60^\circ$ (these simulations will be referred to as \texttt{R75\_ang20}, and \texttt{R75\_ang60}; all the other parameters except for the inclination angle are the same as in \texttt{R75\_ang0}). In the first series of runs we keep $\sigma^{\rm LC}\approx 500\text{-}1000$ and vary $\gamma_{\rm rad}^{\rm LC}$ to capture the following regimes: $\gamma_{\rm rad}^{\rm LC}/\sigma^{\rm LC}=1/15,~1/3,~2,~\infty$ (where $\gamma_{\rm rad}=\infty$ corresponds to simulation without synchrotron cooling; for future reference, these simulations we denote as \texttt{R75\_ang20\_gr1o15}, \texttt{R75\_ang60\_gr1o15}, ..., \texttt{R75\_ang60\_grINF}).


\begin{figure*}[htb]
\centering
\includegraphics[width=1.5\columnwidth]{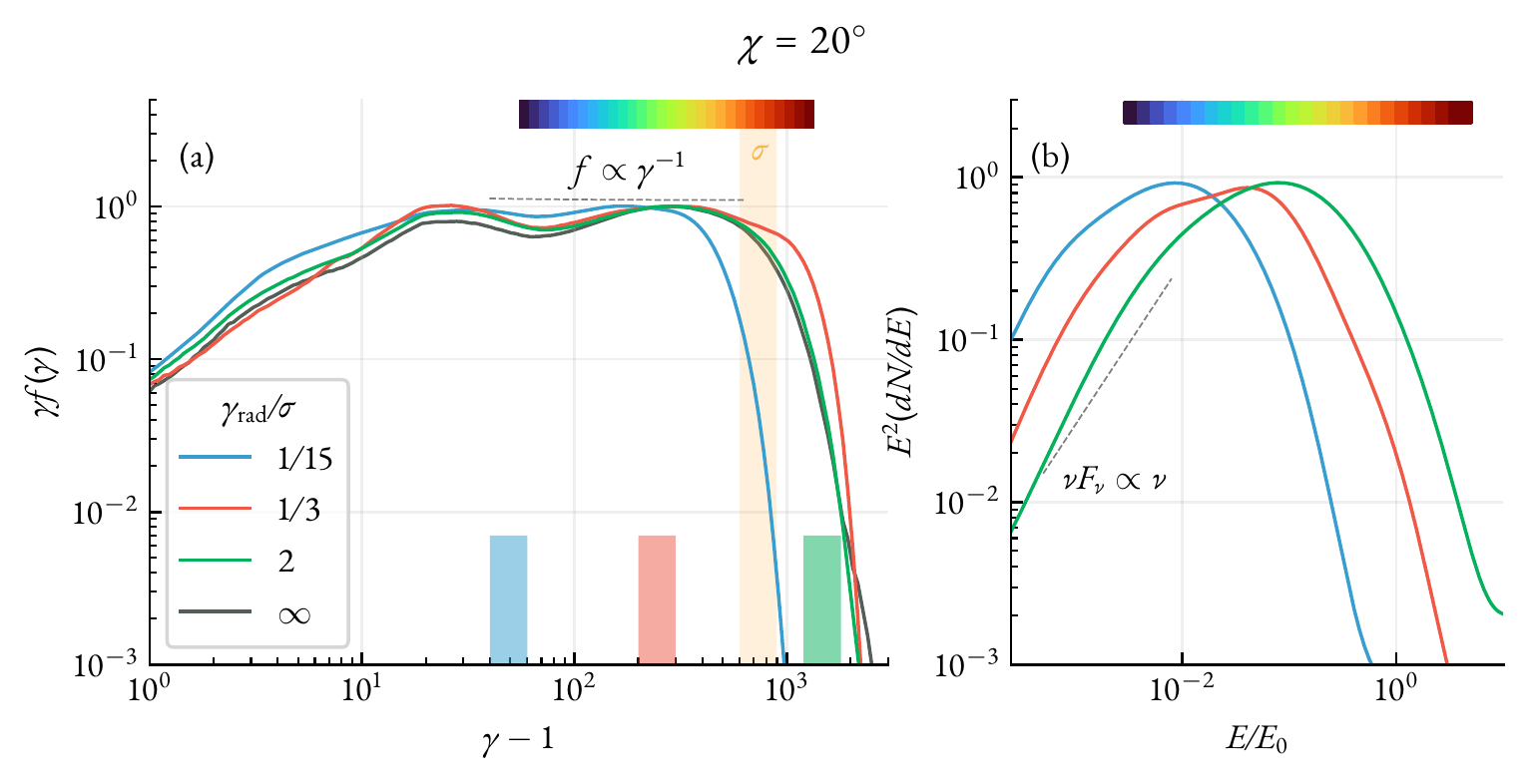}
\includegraphics[width=1.5\columnwidth]{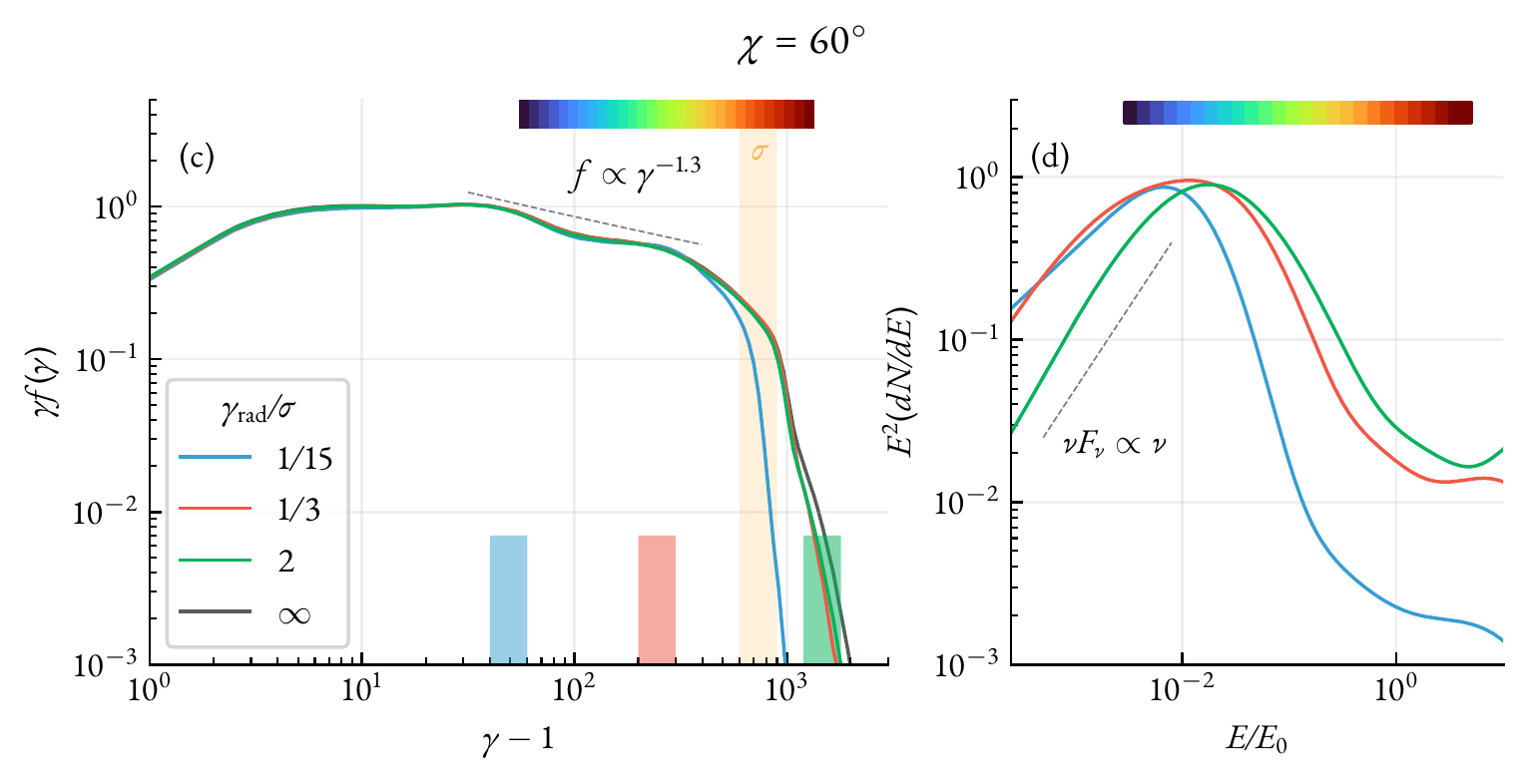}
\caption{
(a, c) particle and (b, d) photon spectra for the \texttt{R75\_ang20}, and \texttt{R75\_ang60} simulations in different synchrotron cooling regimes. Effective $\sigma$ at the light cylinder is marked with yellow stripes. Three smaller bars of different colors in (a) and (c) indicate the effective $\gamma_{\rm rad}$. Colorbar at the top of both panels puts particle energies into correspondence with synchrotron peak energies, $E \propto \gamma^2 B_{\rm LC}$. Photon energies are normalized to $E_0\propto \left(\sigma^{\rm LC}\right)^2 B_{\rm LC}$. While particle spectra look almost identical (except for the strongest cooled case), peaks of photons are shifted to smaller energies for smaller $\gamma_{\rm rad}/\sigma$. Only particles in the current layer are accounted for. Simulations without synchrotron cooling (black lines) are not shown in panels (b) and (d), since we do not collect photons in that case.
}
\label{fig:psr-spec20}
\end{figure*}



Particle distribution functions in the current sheet are also very similar for all the values of the cooling strength, as shown in Figures~\ref{fig:psr-spec20}a,c. In all of the studied cases particles in the layer are able to accelerate to $\gamma\sim\sigma^{\rm LC}$ in x-points, forming a power law of $f\propto \gamma^{-1}\text{-}\gamma^{-2}$ (for larger values of $\chi=60^\circ$ the spectrum steepens from $\gamma^{-1}$ to about $\gamma^{-1.3}$). Only in the most strongly cooled simulation we see that the cutoff of particle distribution is slightly shifted towards lower energies. This marginal difference is likely an effect of comparably small scale separation in our simulations, and will most likely be unnoticable for realistic systems. In weakly cooled simulations above $\gamma>\sigma^{\rm LC}$ we see a smooth transition to $\gamma^{-2.5}\text{-}\gamma^{-3}$ and, eventually, to an almost exponential cutoff. Naively, one would expect that in the reconnection process with strong cooling particles cannot gain energies higher than $\gamma_{\rm rad} m_e c^2$, as for these particles the cooling becomes faster than the acceleration. As mentioned above, this is not necessarily true, as the acceleration takes place in the region of the current sheet where there is virtually no cooling (this has also been demonstrated in 2D simulations by \citealt{2014ApJ...782..104C}, \citealt{2016ApJ...826..221K}, \citealt{2019ApJ...877...53H}, etc). On the other hand, in the strong cooling regime once particles leave the accelerating regions (either back to the upstream or into plasmoids), they very quickly lose their energies without a chance to get reaccelerated again to energies larger than a few $\sigma m_e c^2$. {As a result, the non-thermal distribution of particles, even in the case of a very strong cooling, is determined by the acceleration at x-points and extends up to $\gamma\sim \text{few }\sigma$ \citep[see, e.g.,][]{2022PhRvL.128n5102S}.}


In Figures \ref{fig:psr-spec20}b,d we show the spectra of emitted synchrotron photons for both $\chi=20^\circ$ and $\chi=60^\circ$ simulations {for different cooling strengths}. In all cases we see a recurring pattern in photon spectra: a rise $\nu F_\nu\propto \nu$, a transition with a peak and a decay at higher energies. Power-law index for the distribution of photons is roughly consistent with the power-law index for the distribution of particles, i.e., $f\propto \gamma^{-p}$ leads to $\nu F_\nu \propto \nu^{-(p-3)/2}$.


Since the majority of the particles are unable to retain energies $\gamma > \gamma_{\rm rad}$ (and since $\sigma > \gamma_{\rm rad}$), the spectral peak in photon energies, $E_{\rm p}$, roughly corresponds to the synchrotron emission of particles with $\gamma\sim \gamma_{\rm rad}$: $E_{\rm p} \approx \hbar \omega_B^{\rm LC}(\gamma_{\rm rad}^{\rm LC})^2$. This is evident from a comparison of the colorbars in \ref{fig:psr-spec20}a,b: we see a shift of the emission peak towards lower energies for simulations with stronger cooling. The spectrum, nevertheless, extends further, as particles are still able to accelerate to energies $>\gamma_{\rm rad}$ in the x-points, albeit rapidly radiating away. In the case of moderate-to-weak cooling ($\sigma<\gamma_{\rm rad}$), the majority of particles is accelerated to $\gamma\sim\sigma$, and the peak is controlled by the value of $\sigma$: $E_{\rm p} = \hbar \omega_B^{\rm LC}(\sigma_{\rm}^{\rm LC})^2 $ (see green lines in Figures~\ref{fig:psr-spec20}b,d).

Note that we do not capture significant intermittency at higher energies for the strong cooling simulation, which was prominent in 2D localized simulations by \cite{2019ApJ...877...53H}. {Because of this, simulations with stronger cooling also have smaller spectral cutoff energy together with a smaller spectral peak, despite the fact that the cutoff in particle distribution is unaffected.} In 2D the high energy intermittency, which was responsible for extending the cutoff to values close to $\hbar \omega_B \sigma^2$, was primarily caused by merger events of plasmoids of different sizes varying from just a few to hundreds of plasma skin depths. In our global 3D simulations plasmoids are at most several skin depths in size, and there is only a handful of them in the current sheet. This striking difference in the separation of scales between local 2D and global 3D simulations, as well as the overall complexity in the structure of the current sheet in 3D, may explain the observed difference. {Future large scale localized 3D simulations of magnetic reconnection in the high-$\sigma$ strong cooling regime will be able to clarify this conflict and provide a more definitive picture.}

\begin{figure}[htb]
\centering
\includegraphics[width=0.75\columnwidth,trim=20 0 10 0,clip]{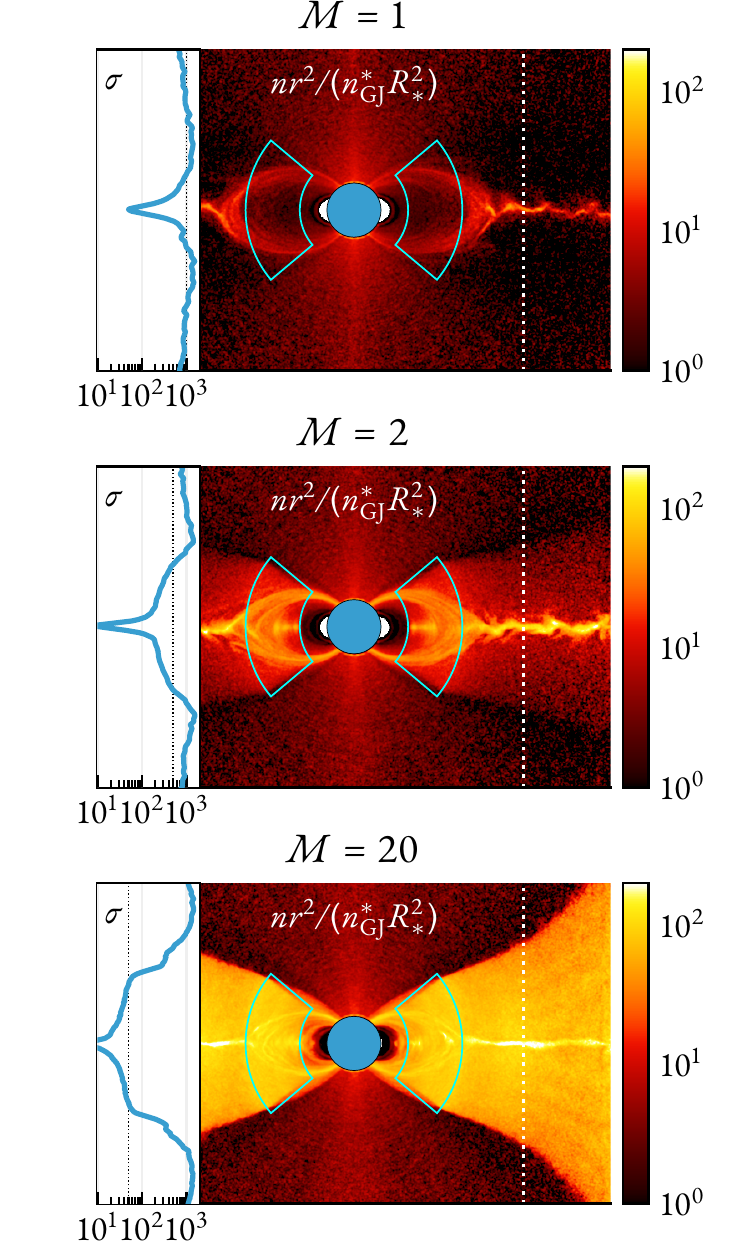}
\caption{\NEW{Plasma density, compensated for the falloff with the distance, $nr^2/(n_{\rm GJ}^*R_*^2)$,} for three simulations \texttt{R75\_ang0} with different pair loading rates. Additional injection region is highlighted with cyan. Newly created particles are initialized with zero velocity. From top to bottom, $M=1$ (no extra injection, all particles originate at the surface), $M=2$ (one particle is injected per each surface-injected particle), $M=20$. To the left of each panel we plot the magnetization of the steady-state solution for each case as a function of vertical coordinate (corresponding slice is shown with a white dashed line).
}
\label{fig:psr-denspp}
\end{figure}

\subsection{Effects of additional pair-loading}
\label{sec:pairloading}

To test how particle acceleration and radiation spectra change with the value of magnetization, $\sigma^{\rm LC}$, we artificially load the current sheet with extra plasma to alter the effective value of $\sigma$ in the current layer (see Figure~\ref{fig:psr-denspp}). For each pair injected from the surface we initialize $M - 1$ additional pairs at rest in the cyan region shown in Figure~\ref{fig:psr-denspp}.\footnote{For computational purposes we limit injection only to fieldlines close to the equator that participate in reconnection. Since particles are well magnetized and any transport in the transverse direction to the magnetic field is suppressed, the fieldlines at higher altitudes have little-to-no effect on the reconnection process.} All the other parameters, such as $\gamma_{\rm rad}^{\rm LC}\approx 10^3$ and $V_{\rm pc}\approx 5000~m_e c^2$, are kept constant. These ``secondary'' pairs rapidly catch up with the bulk $\bm{E}\times\bm{B}$ outflow, and their ultimate effect is the decrease of the effective magnetization $\sim M$ times (as shown in the left panels of Figure~\ref{fig:psr-denspp}).

\begin{figure*}[htb]
\centering
\includegraphics[width=1.75\columnwidth]{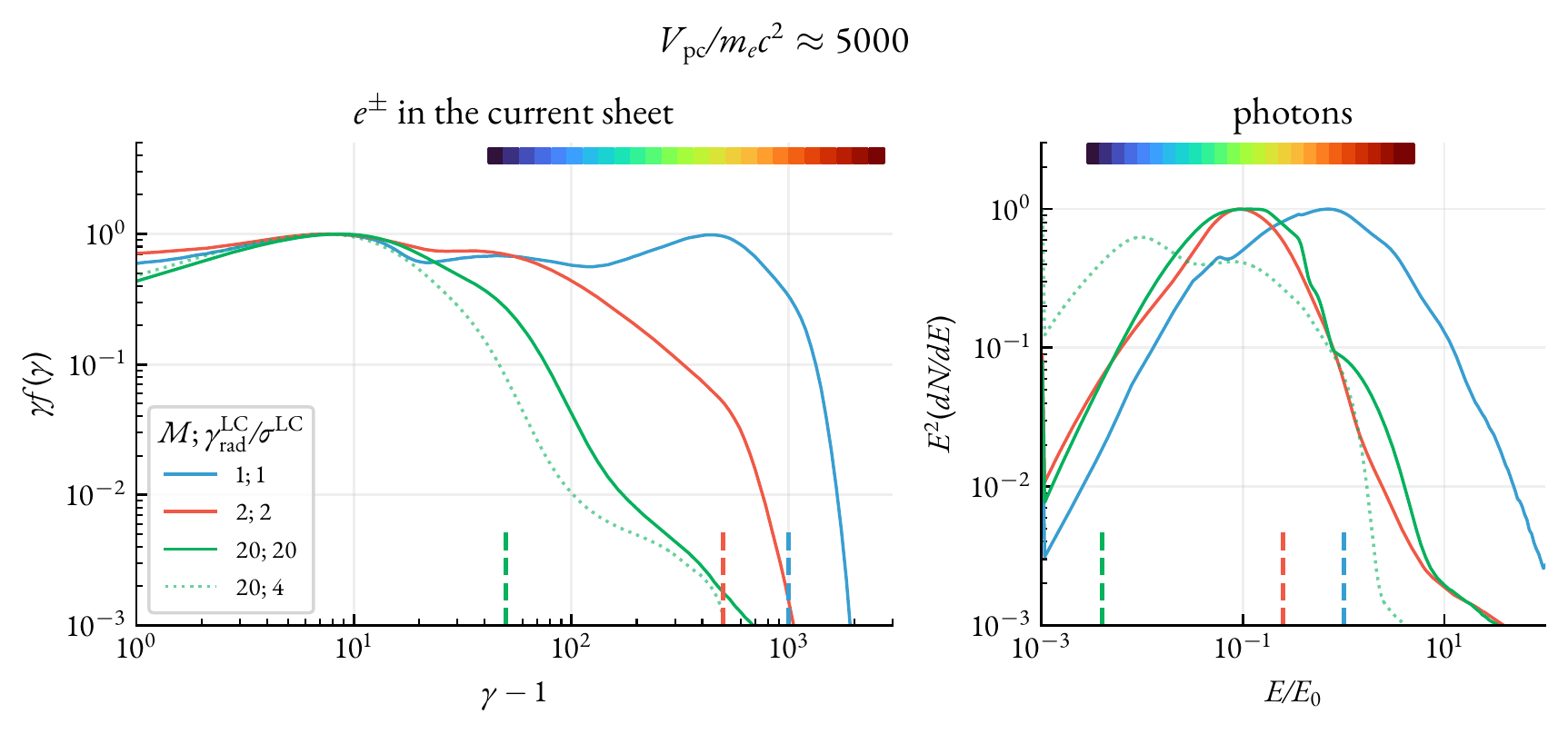}
\caption{Particle energy distributions in the current layer and photon spectra for simulations shown in Figure~\ref{fig:psr-denspp}. Dashed lines indicate the effective magnetization near the light cylinder, $\sigma^{\rm LC}$, and the corresponding photon energies, $E\propto \left(\sigma^{\rm LC}\right)^2 B_{\rm LC}$. We also show an additional simulation (with a dotted line) where we enhance cooling for the $M=20$ case by decreasing $\gamma_{\rm rad}\approx 200$. In all the other cases the cooling strength is fixed at $\gamma_{\rm rad}\approx 1000$. Photon energies are normalized to $E_0\propto \left(\sigma^{\rm LC}_{\mathcal{M}=1}\right)^2 B_{\rm LC}$.}
\label{fig:psr-specpp}
\end{figure*}

In Figure~\ref{fig:psr-specpp} we show corresponding particle distributions in the current layer, as well as the emerging radiation spectra for all these simulations. The blue line shows the fiducial case without additional injection. In that simulation the magnetization is $\sigma^{\rm LC}\sim 10^3$, and particles in the current layer form an extended power-law of $f\propto\gamma^{-1}$ extending to $\gamma\sim \sigma^{\rm LC}$. In this moderate cooling regime the peak of radiation spectrum corresponds roughly to the synchrotron peak of the particles with $\gamma\sim\gamma_{\rm rad}^{\rm LC}\sim 10^3$. With increasing pair density, $M$, magnetization at the light cylinder drops proportionally (as shown with red and green lines in the particle spectra of Figure~\ref{fig:psr-specpp}). In cases of $M=2$, and $M=20$, the magnetization values near the Y-point are, respectively, $\sigma^{\rm LC}\approx500$, and $\approx 50$. Particles form a clear hard power-law distribution up to $\gamma\sim \sigma^{\rm LC}$ with a steepening to $f\propto\gamma^{-2}\text{-}\gamma^{-3}$ at higher energies, $\gamma>\sigma^{\rm LC}$, followed by a cutoff. In these cases pairs are still able to accelerate to energies $>\sigma^{\rm LC}m_e c^2$ since the cooling is weak ($\gamma_{\rm rad}^{\rm LC}\sim 10^3 < \sigma^{\rm LC}$). In the simulation shown with a dashed green line we inhibit this secondary acceleration by enhancing the synchrotron cooling strength (decreasing $\gamma_{\rm rad}^{\rm LC}$ to $\approx 200$). These simulations clearly demonstrate that the particle acceleration potential of the pulsar outer magnetosphere is indeed determined by the local magnetization parameter, $\gamma_{\rm max}\sim \sigma^{\rm LC}$, rather than the electric potential drop near the polar cap, $V_{\rm pc}/m_e c^2$, defined in equation~Eq.~\eqref{eq:psr-vpc}. In pulsars these values are correlated, with $\sigma^{\rm LC}$ being significantly smaller:

\begin{equation}
    \sigma^{\rm LC}\approx 
    \frac{1}{2}
    \left(
        \frac{n^{\rm LC}}{n_{\rm GJ}^{\rm LC}}
    \right)^{-1}
        \frac{V_{\rm pc}}{m_e c^2} \ll \frac{V_{\rm pc}}{m_e c^2}.
\end{equation}
where $n^{\rm LC} / n^{\rm LC}_{\rm GJ}\gg 1$ is the plasma multiplicity near the light cylinder.

Photon spectra in these runs are consistent with our earlier predictions (see Sec.~\ref{sec:part_photon_dist}). In the first run (blue line) the spectral peak is set by particles with energies $\gamma\sim\gamma_{\rm rad}^{\rm LC}\sim \sigma^{\rm LC}$. When we decrease $\sigma^{\rm LC}$ (while keeping the cooling strength, $\gamma_{\rm rad}^{\rm LC}$, constant) we see that the peak is only marginally sensitive to $\sigma^{\rm LC}$ (which drops from $10^3$ in $M=1$, to $200$ in $M=2$, and to $50$ in $M=20$), and is controlled by particles with $\gamma\sim\gamma_{\rm rad}^{\rm LC}$ (hence the similarity of the red and green curves). When we decrease $\gamma_{\rm rad}^{\rm LC}$ to $200$ for the $\sigma^{\rm LC}\sim 50$ run (green dashed line), the peak is shifted towards lower energies (despite the fact that particle distributions are only marginally affected).

\section{Discussion}
\label{sec:psr-observ}
\subsection{Summary}

In this work we closely inspect 3D PIC simulations of neutron star magnetospheres with the strong synchrotron radiation reaction force included. Our primary focus is the energy dissipation and the plasma dynamics in the reconnecting current sheet. To be able to properly capture the kinetic physics governing the energy extraction in the current layer, we push the separation of macro-to-micro scales in our simulations to $\sim 200$.

We show that the rate of plasma inflow into the reconnecting equatorial current sheet, which is controlled by plasma kinetics at the scales of microscopic x-lines, determines the dissipation rate in the entire magnetosphere. This puts a stringent constraint on how much electromagnetic energy can be dissipated (and, ultimately, radiated) in the current layer within a few light cylinders. From simulations we find that this dissipation rate is almost insensitive to both the bulk motion of the background unreconnected plasma, and the synchrotron cooling strength, as long as the magnetization is high: $\sigma^{\rm LC}\gg 1$. The fraction of the dissipated energy within the region between $R_{\rm LC} < r < 2R_{\rm LC}$ in the outer magnetosphere varies between $1\text{-}10\%$, with the exact value depending only on the pulsar inclination angle.

Due to the fast synchrotron cooling timescale, reconnection in the equatorial current layer proceeds in the radiatively efficient regime. Large fraction of the dissipated electromagnetic energy is radiated away by the accelerated pairs. Since cooling is always faster than the dynamical time of the current sheet, the amount of radiated energy is insensitive to the cooling strength and only depends on the amount of magnetic energy dissipated. 

We confirm that the particle distribution is almost unaffected by synchrotron cooling, with particles in the current sheet forming a hard power-law spectrum with steep cutoff at energies of $\sim \sigma^{\rm LC}m_e c^2$. Photon spectra, on the other hand, are evidently different. In the marginal cooling regime, $\gamma_{\rm rad}^{\rm LC}\sim \sigma^{\rm LC}$, both the peak and the high-energy cutoff of the emission are controlled by the magnetization parameter, $E_{\rm p}\approx E_{\rm cut}\approx \hbar\omega_{B}^{\rm LC}(\text{few}\cdot\sigma^{\rm LC})^2$. In the magnetospheres where synchrotron cooling is strong, $\gamma_{\rm rad}^{\rm LC} < \sigma^{\rm LC}$, plasmoids that carry the bulk of the plasma and contribute to the high energy emission the most quickly cool down to the characteristic temperatures of $T_e \sim \gamma_{\rm rad}^{\rm LC}m_e c^2$ \citep[see, e.g.,][]{2014ApJ...780....3U}. Our simulations suggest that the peaks of the emission are thus controlled by the synchrotron emission of particles with energies comparable to $T_e$: $E_{\rm p} \approx \hbar \omega_B^{\rm LC}(T_e/m_e c^2)^2\sim \hbar \omega_B^{\rm LC}(\gamma_{\rm rad}^{\rm LC})^2\approx 20$ MeV. 

\subsection{Observational Implications}
\label{sec:observ}

Observations by the \emph{Fermi} telescope during the last decade have shown that some of the pulsars with $\dot{E}\gtrsim10^{34}$ erg/s emit in $\gamma$-rays with characteristic luminosities ranging between $0.1\%$ and $10\%$ of the spin-down power. Our 3D PIC simulations provide strong evidence that this emission is powered by magnetic reconnection in the equatorial current sheets of pulsar magnetospheres. Reconnection occurs in the non-linear stage of the tearing instability which develops in the equatorial current sheet on microscopic timescales, $\omega_{\rm t}^{-1}\approx \Gamma w_{\rm cs}/c$, where $w_{\rm cs}$ is the characteristic width of the sheet, and $\Gamma\approx\sqrt{R_{\rm LC}/R_*}$ is the bulk Lorentz factor of the upstream plasma in the wind.\footnote{The growth rate of the instability is defined as $c/w_{\rm cs}$ in the rest frame of the upstream plasma and should be Lorentz-boosted appropriately.} The width of the current sheet, $w_{\rm cs}$, is defined by the pressure equilibrium, and is set by the Larmor radii of particles with characteristic Lorentz factors of $\gamma\approx (T_e/m_e c^2)\approx \gamma_{\rm rad}^{\rm LC}$: $w_{\rm cs}\approx \gamma_{\rm rad}^{\rm LC} m_e c^2/(|e| B_{\rm LC})$, where the $\gamma_{\rm rad}^{\rm LC}$ parameter is set by radiation reaction force. In this paper we explicitly assumed that the plasma microphysics is disentangled from the large scale structure of the pulsar magnetosphere by taking a large-enough separation of micro-to-macro scales. For this assumption to be valid it is necessary for the tearing instability to develop much faster than the dynamical time of the magnetosphere -- the pulsar rotation period, $\Omega^{-1}$. The ratio of these timescales is always large for $\gamma$-ray pulsars, $\omega_{\rm t}/\Omega\approx 10^3\dot{\mathcal{E}}_{36}/\sqrt{\mathcal{B}_{12}}$, meaning that at least for the young pulsars with $\dot{E}\gtrsim 10^{35}$ erg/s the reconnection develops almost instantly in the current sheet near the Y-point.\footnote{Here and further in this section we normalize all the quantities to fiducial values of $\dot{\mathcal{E}}_{36}\equiv \dot{E}/(10^{36}\text{erg/s})$, $\mathcal{B}_{12}\equiv B_*/(10^{12}\text{G})$, and $\mathcal{K}_4\equiv \kappa/ 10^4$ (plasma multiplicity w.r.t. the local Goldreich-Julian value). For reference, $\dot{E}$ varies between $10^{32}$ and $10^{38}$ erg/s (Vela: $7\times 10^{36}$ erg/s, Crab: $4\times 10^{38}$ erg/s), $B_*$ weakly changes between $10^{12}\text{-}10^{13}$ G, and $\kappa\approx 10^2\text{-}10^7$ (with $\gtrsim 10^4$ being the case only for a few youngest pulsars, e.g., the Crab). In practice $\kappa$ is determined by the dynamics of the polar cap discharge, and the pair-production efficiency near the light cylinder, both of which indirectly depend on $\dot{E}$ and $B_*$ (or $P$ and $\dot{P}$). But for the purposes of this discussion we omit this dependency.}

The amount of energy dissipation within a few light cylinders beyond the Y-point, where the high-energy radiation originates, is determined by the reconnection rate, $\beta_{\rm rec}\approx 0.1$, and the pulsar inclination angle. Importantly, we find that this rate is not suppressed by the presence of the bulk flow of the upstream plasma along the magnetic field lines, as long as the Lorentz factor of the Alfv\'en velocity is larger than that of the bulk flow, $\sqrt{\sigma^{\rm LC}}\gg \Gamma\approx\sqrt{R_{\rm LC}/R_*}$.\footnote{The value of $\Gamma$, which is the Lorentz factor of the bulk motion of pairs flowing from the polar cap to the outer magnetosphere, is set by the last curvature photons to pair produce before escaping the polar cap discharge region. The energy of pairs produced by these photons is determined by the angle, $\psi$, between the photon momentum and the magnetic field: $\gamma_\pm \approx 1/\psi$ (this is valid in the limit where the energies of photons are large: $\gg m_e c^2$, and the angle $\psi\ll 1$; see, e.g., \citealt{2010MNRAS.408.2092T}). Since the discharge operates up to $\approx R_*$ from the surface of the star, the last photons to pair produce will have $\psi\approx R_*/R_c$, where $R_c$ is the radius of curvature of the magnetic field lines $R_c\approx R_*\sqrt{R_{\rm LC}/R_*}$. From these we finally find: $\Gamma = \gamma_\pm^{\rm (last)} \approx \sqrt{R_{\rm LC}/R_*}$.} In young pulsars $\Gamma\approx 20~\mathcal{B}_{12}^{1/4}/\dot{\mathcal{E}}_{36}^{1/8}$, whereas $\sqrt{\sigma^{\rm LC}}\approx 500~ \dot{\mathcal{E}}_{36}^{1/4}\mathcal{K}_4^{-1/2}$, implying that the two are separated by at least an order of magnitude. However, even when $\sqrt{\sigma^{\rm LC}}\sim \Gamma$ (which might be the case for the older pulsars with $\dot{E}\approx 10^{34}$ erg/s) our findings indicate that the rate is only marginally slower (by a factor of two at most), which is virtually unnoticeable for the purposes of observations.

Our simulations demonstrate that the dissipated power is further radiated away via synchrotron radiation, which we identify as the main emission mechanism in our simulations. Curvature radiation in this context is negligible \citep[contrary to what has been reported by, e.g.,][]{2018ApJ...857...44K}, as the dynamically strong cooling prevents particles from accelerating to energies beyond $\sigma^{\rm LC}m_e c^2$, and the Larmor radii of the most energetic particles, $\rho_L$, are limited to microscopic plasma scales, $R_{\rm LC}/\rho_L\approx 2\cdot 10^4~\mathcal{K}_4\gg 1$. For the Vela-like pulsars, where the discharge near the polar cap is the main source of pairs, and there is no additional pair production in the current sheet, most of the dissipated energy is emitted directly in the form of the highest-energy synchrotron photons around one-to-few GeV, leading to characteristically high values of $L_\gamma/\dot{E}\approx 1\%\text{-}10\%$. The youngest Crab-like pulsars, on the other hand, produce sufficient amount of high-energy photons, so that pair creation near the light cylinder becomes an important source of additional pair  plasma \citep{1996A&A...311..172L, 2019ApJ...877...53H}. To estimate the fraction of the dissipated magnetic energy that goes into these secondary pairs we need to evaluate the effective optical depth for the $\gamma$-ray photons, $\tau_{\gamma x} = f_{\gamma x}\sigma_T n_x R_{\rm LC}$, where $f_{\gamma x}\approx 0.25$ (given by the maximum cross section of the pair-production process: $\sigma_{\gamma\gamma}^{\rm (max)}\approx 0.25 \sigma_T$; see, e.g., \citealt{1985quel.book.....A}), and $n_x$ is the number density of X-ray photons, the presence of which is required for the GeV photons to pair produce.\footnote{See also similar arguments in the context of the M87* by \citealt{Ripperda2022}.} The effective number density of the low-energy photons can be estimated as $n_x\approx (L_x/c\varepsilon_x)/(4\pi R_{\rm LC}^2)$, where $L_x$ is the luminosity in X-rays, and $\varepsilon_x$ is the characteristic energy of X-ray photons. Pair production cross section is the largest for the photons that satisfy $\varepsilon_x\varepsilon_\gamma \approx (m_e c^2)^2$ (here, $\varepsilon_\gamma\approx$ GeV). Substituting all the values, and taking $L_x/\dot{E}\sim 1\%$, which is the case for the youngest pulsars \citep{2011ApJ...733...82M}, we finally find: $\tau_{\gamma x}\approx 10^{-2}\dot{\mathcal{E}}_{36}^{5/4}/\sqrt{\mathcal{B}_{12}}$. For the Crab, $\dot{E}\approx 4\cdot 10^{38}$ erg/s, and we find that $\tau_{\gamma x}\approx 1$. The relatively large value for the optical depth means that the non-negligible fraction of the dissipated energy is deposited into secondary pairs produced in the reconnection upstream. Characteristic energies of the produced pairs can be estimated as $\gamma_{\rm sec}\approx E_\gamma / (m_e c^2)$, where $E_\gamma\approx 0.1\text{-}1$ GeV is the charactertic energy of the pair-producing photons. For the Crab, $\gamma_{\rm sec}\approx 10^3$ and the synchrotron emission of these secondary pairs corresponds to energies $E_{\rm sec}\approx \hbar \gamma_{\rm sec}^2 |e| B_{\rm LC}/(m_e c)\approx 1$ keV.




The maximum energy to which particles accelerate during the reconnection process is controlled by the plasma magnetization parameter near the light cylinder, few-to-ten $\sigma^{\rm LC}$, which in turn determines the cutoff energy in the observed emission. The synchrotron cutoff can be expressed as $E_{\rm cut}\approx (3/2)\hbar \gamma^2_{\rm cut} |e|\tilde{B}_\perp/(m_e c)$, where $\tilde{B}_\perp^2=(\bm{E}+\bm{\beta}\times \bm{B})^2-(\bm{\beta}\cdot\bm{E})^2$ \citep[see, e.g.,][]{2016MNRAS.457.2401C}, and $\gamma_{\rm cut}\approx \text{few}\cdot \sigma^{\rm LC}$. For reconnection in the strong cooling regime most of the dissipated energy is radiated away via emission from highest-energy pairs. The dissipation power is equal to $|e| E_{\rm rec} c$, where $E_{\rm rec}\approx \beta_{\rm rec} B_{\rm LC}$, while the synchrotron radiation power is $\sigma_T c \gamma_{\rm cut}^2 \tilde{B}_\perp^2/(4\pi)$. From the definition of $\gamma_{\rm rad}^{\rm LC}$ we may then find: $\gamma_{\rm cut}\tilde{B}_\perp \approx \gamma_{\rm rad}^{\rm LC}B_{\rm LC}$, where $B_{\rm LC}$ is the upstream magnetic field strength. Substituting this into the expression for the cutoff energy yields: $E_{\rm cut}\approx (3/2)\hbar \left(\gamma_{\rm rad}^{\rm LC}\right)^2|e|B_{\rm LC}/(m_e c)\left(\gamma_{\rm cut}/\gamma_{\rm rad}^{\rm LC}\right)\approx (20~\text{MeV})\left(\gamma_{\rm cut}/\gamma_{\rm rad}^{\rm LC}\right)$. The cutoff energy is then uniquely set by the ratio $\gamma_{\rm cut}/\gamma_{\rm rad}^{\rm LC}$, resulting in $E_{\rm cut}\approx (200~\text{MeV})\dot{\mathcal{E}}_{36}^{7/8}\mathcal{B}_{12}^{-1/4}/\mathcal{K}_4$. For Vela-like pulsars, $\kappa\approx 10^3\text{-}10^4$ \citep{2015ApJ...810..144T}, $\dot{E}\approx 10^{36}\text{-}10^{37}$ erg/s, resulting in $E_{\rm cut}\approx 1\text{-}10$ GeV.\footnote{For the least energetic $\gamma$-ray-emitting pulsars with $\dot{E}\lesssim 10^{36}$ erg/s it is likely that the polar cap discharge is less efficient, and the resulting plasma multiplicity near the light cylinder is $\kappa \lesssim 10^3$, which yields the same cutoff energy.} 

For the most energetic Crab-like pulsars with a strong magnetic field near the light cylinder ($B_{\rm LC}\sim 4\cdot 10^6$ G for Crab) pair creation near the Y-point and beyond is important. The amount of produced pairs is proportional to the optical depth, $\tau_{\gamma x}$, estimated earlier, and the number of GeV photons. The production rate can be estimated as $\dot{N}_\pm\approx \tau_{\gamma x}L_\gamma/\varepsilon_\gamma$, where $\varepsilon_\gamma\approx$ GeV. To obtain the multiplicity $\kappa$ we compare this with the GJ flux, $\dot{N}_{\rm GJ}\approx \sqrt{c \dot{E}}/|e|$, where we assumed that the $\dot{N}_{\rm GJ}e$ is equal to the GJ current: $\sqrt{c\dot{E}}$. We then find the multiplicity of pair-production near the light cylinder $\kappa_{\rm LC}=\dot{N}_\pm/\dot{N}_{\rm GJ}\approx 40~\dot{\mathcal{E}}_{36}^{7/4}\mathcal{B}_{12}^{-1/2}(L_x/1\%\dot{E})(L_\gamma/0.1\%\dot{E})$. For the Crab pulsar, $L_x/\dot{E}\approx 1\%$, and $L_\gamma/\dot{E}\approx0.1\%$ yielding $\kappa_{\rm LC}\approx 10^6$. For the Vela, on the other hand, $L_x/\dot{E}\approx 0.01\%$, and $L_\gamma/\dot{E}\approx 1\%$ yielding $\kappa_{\rm LC}\approx 3$ (similar to what has been obtained by \cite{1996A&A...311..172L}.\footnote{Notice, that the total multiplicity $\kappa = \kappa_{\rm LC} + \kappa_{\rm PC}$, where $\kappa_{\rm PC}$ is the multiplicity of the pair cascade near the polar cap.} Substituting $\kappa\approx \kappa_{\rm LC}$ for the Crab to the relation for the cutoff energy we then find $E_{\rm cut}\approx 0.1\text{-}1$ GeV.

To reiterate, while the amount of the dissipated magnetic energy is uniquely set by the microphysics of the reconnection and the inclination angle, the luminosity of the observed emission can be somewhat different between the high-$\dot{E}$ Crab-like, and the low-$\dot{E}$ Vela-like pulsars. The cutoff energies are dictated by the highest-energy pairs, and in both cases the numbers are close to the observed one-to-few GeV. Spectral peaks, on the other hand, are different. In the case of Vela-like pulsars most of the dissipated Poynting flux is radiated away via synchrotron at energies close to the cutoff around $1\text{-}10$ GeV. For the Crab-like pulsars, however, secondary pair production near the light cylinder is important, and a large fraction of the dissipated energy goes into pairs which further re-radiate via synchrotron at a range of energies spanning all the way to keV, resulting in a much broader spectrum. This yields a smaller $\gamma$-ray luminosity in the \emph{Fermi} band (between $0.1$ and $100$ GeV), which is clearly observed by \citealt{2013ApJS..208...17A} where authors report $\dot{E}\propto L_\gamma^{1/2}$ beyond $\dot{E}\gtrsim 10^{36}$ erg/s.

\subsection{Future work}
In our simulations we did not include self-consistent pair production, either the $\gamma B\to e^\pm$ process near the polar cap, or the $\gamma\gamma\to e^\pm$ (Breit-Wheeler) process near the light cylinder, instead choosing to provide a constant inflow of plasma from the surface of the star. Our reasonable assumption was that as long as enough plasma is provided to the magnetosphere ($n\gtrsim n_{\rm GJ}$), its general structure, as well as the dynamics of the current layer would not depend on how exactly the plasma was supplied. However, as mentioned earlier, for the youngest Crab-like pulsars the secondary pairs produced by the two-photon pair production in the outer magnetosphere can carry a large fraction of the dissipated energy, re-radiating it at optical to X-ray frequencies. Additionally, as was shown in 2D simulations of an isolated current sheet \citep{2019ApJ...877...53H}, and as mentioned in section~\ref{sec:pairloading} of this paper, these secondary pairs inhibit the acceleration process during reconnection, thus controlling the extent of the gamma-ray signal. 
To understand this process in more details, and also study the intermittency, the light curves, and the polarization characteristics of the distinct X-ray signal these pairs produce, global 3D simulations with proper two-photon pair production physics are necessary. \texttt{Tristan-MP v2} code \citep{tristanv2}, used in this paper, supports the capabilities of tracking individual photons and modeling their interaction pair-by-pair. In the future we plan to apply this capability, coupled with the hybrid guiding center particle pusher, to model the pair-production process and its dynamics in the global context of the magnetosphere more realistically. While we think our main conclusions outlined in section~\ref{sec:observ} will still hold, these more robust simulations will allow to more self-consistently predict the optical/X-ray luminosity for the young pulsars (which we simply postulated empirically for the purposes of our estimations), to understand the long term evolution of the secondary pairs, as well as their effect on the structure of the pulsar wind.

\section*{Acknowledgements}
Authors would like to thank Andrei Beloborodov and Ioannis Contopoulos for insightful comments and thorough remarks\NEW{, as well as the anonymous reviewer for concise and helpful comments}. This work was partially supported by the U.S. Department of Energy under contract number DE-AC02-09CH11466. The United States Government retains a non-exclusive, paid-up, irrevocable, world-wide license to publish or reproduce the published form of this manuscript, or allow others to do so, for United States Government purposes. This work was supported by NASA grant 80NSSC18K1099 and NSF grants AST-1814708 and PHY-1804048. AS is supported by Simons Investigator grant 267233. Computing resources were provided and supported by Princeton Institute for Computational Science and Engineering and Princeton Research Computing. This research was facilitated by Multimessenger Plasma Physics Center (MPPC), NSF grant PHY-2206607.

\vspace{2cm}

\appendix

\section{Numerical setup and limitations}
\label{sec:psr-appendixA}

In this Appendix we present some dimensionless relations both in the context of astrophysical pulsars, and our simulations. We discuss how each of these parameters is up/down scaled in our global simulations relative to realistic values.

Consider a rotating perfectly conducting neutron star with a radius $R_*$ and a magnetic moment $\mu=B_* R_*^3$, {where $B_*$ is the magnetic field strength near the equator}. For simplicity let us consider $\chi=0^\circ$, i.e., the magnetic moment is aligned with the rotation axis. If the magnetosphere is filled with enough plasma to provide the necessary charge density for screening the parallel electric field, then magnetic field lines that originate inside a circular region of radius $R_{\rm pc}$ around the magnetic axis will be open, i.e., will continue to infinity. This region, the \emph{polar cap}, has a radius of $R_{\rm pc} = R_*\sqrt{R_*/R_{\rm LC}}$, where $R_{\rm LC} = c/\Omega$ is the light cylinder of the neutron star. Electric field generated at the surface of the star across the polar cap generates a potential drop, $V_{\rm pc}$, which can be written in a dimensionless form
\begin{equation}
\label{eq:psr-vpc}
    \frac{V_{\rm pc}}{m_e c^2} = \frac{\omega_{B_*}}{\Omega}\left(\frac{R_*}{R_{\rm LC}}\right)^3\approx 2.6\cdot 10^{9}\left(\frac{B_*}{10^{12}~\textrm{G}}\right)\left(\frac{P}{100~\textrm{ms}}\right)^{-2},
\end{equation}
\noindent where $\omega_{B}^* = |e| B_* / m_e c$ is the nominal gyrofrequency at the surface of the star, and $P$ is the rotation period of the star.

Defining the plasma multiplicity close to the light cylinder, $\lambda$, as $n_{\rm LC} = \lambda \Omega B_{\rm LC} / 2\pi c |e|$, the scale separation (ratio of the light cylinder to the plasma skin depth) in the magnetospheric current layer is given by the following expression:
\begin{equation}
\label{eq:psr-scalesep}
    \frac{R_{\rm LC}}{d_e^{\rm LC}}=\left(2\lambda \frac{V_{\rm pc}}{m_e c^2}\right)^{1/2}\approx 7\times 10^{6}\left(\frac{\lambda}{10^4}\right)^{1/2}\left(\frac{B_*}{10^{12}~\textrm{G}}\right)^{1/2}\left(\frac{P}{100~\textrm{ms}}\right)^{-1}.
\end{equation}

The plasma magnetization near the light cylinder is
\begin{equation}
    \sigma^{\rm LC} = \frac{1}{2\lambda}\frac{V_{\rm pc}}{m_e c^2} \approx 10^5 \left(\frac{\lambda}{10^4}\right)^{-1}\left(\frac{B_*}{10^{12}~\textrm{G}}\right)\left(\frac{P}{100~\textrm{ms}}\right)^{-2}.
\end{equation}

The duration, resolution and other parameters of our simulations are chosen in such a way as to satisfy several requirements, while still being computationally feasible. These requirements are the following:
\begin{enumerate}
    \item $T\gtrsim 3 P$: the durations of our simulations are typically a few rotation periods to ensure the initial transient state has passed and a steady-state is reached;
    \item $L^3\gtrsim (5 R_{\rm LC})^3$: size of the domain has to be large enough to fit at least $1 R_{\rm LC}$ of the current sheet;
    \item $R_*\gg \Delta x$: stellar surface needs to be well resolved;
    \item $V_{\rm pc}/m_e c^2\gg 1$: potential drop at the polar cap needs to be very large;
    \item $\lambda > 1$: to ensure enough plasma is provided to screen the accelerating electric field, the plasma multiplicity has to be at least a few;
    \item $d_e^* \sim \Delta x$: the plasma skin depth at the surface of the star has to be marginally resolved (or at least not strongly underresolved);
    \item $d_e^{\rm LC}\gtrsim \Delta x$: the plasma skin depth near the magnetospheric current sheet has to be resolved to properly capture the reconnection process;
    \item $R_{\rm LC} / d_e^{\rm LC}\gtrsim 100$: the scale separation in the current sheet has to be large;
    \item $\sigma^{\rm LC}\gg 1$: to ensure the reconnection is in high-$\sigma$ regime (as shown in Eq.~\eqref{eq:psr-scalesep} this is guaranteed from the first and second conditions);
    \item $\omega_{B}^{\rm LC}\lesssim \Delta t^{-1}$: to properly capture the synchrotron cooling and gyration of particles in the current layer, their corresponding gyration periods have to be resolved.
\end{enumerate}

In our simulation \texttt{R75\_ang0} we use $L=2040\Delta x$, $R_*=75\Delta x$, $R_{\rm LC}=444\Delta x$ ($P=6200\Delta t$, $\Omega \approx 10^{-3}\Delta t^{-1}$, with the speed of light being $c=0.45\Delta x / \Delta t$), and $\omega_{B}^*\approx 10^3\Delta t^{-1}$.\footnote{This means that gyration of the coldest particles is extremely underresolved near the surface of the star. This is only possible because of the coupled GCA/Boris particle pusher we employ.} This yields $V_{\rm pc}/m_e c^2 \approx 5\times 10^3$, $\omega_{B}^{\rm LC}\sim 5\Delta t^{-1}$ (marginally underresolved for the lowest energy particles). We also typically have $\lambda \sim 5\text{-}10$, and, thus, $d_e^*\approx 0.1\Delta x$, $d_e^{\rm LC}\approx 2\text{-}5\Delta x$, and $R_{\rm LC}/d_e^{\rm LC}\sim 100\text{-}200$. Magnetization, on the other hand, is $\sigma^{\rm LC}\sim 500\text{-}1000$.

\section{Guiding center approximation}
\label{sec:psr-appendixB}

In all of our simulations of pulsar magnetospheres we employ a hybrid particle mover algorithm \citep{2020ApJS..251...10B}. The algorithm allows to switch between two numerical schemes for solving the equation of motion of particles depending on the electric and magnetic field values at particle location, as well as the momentum of the particle. In the GCA (guiding center approximation) regime the motion of the particle is reduced to the motion of its guiding center, ignoring the curvature and $\nabla \bm{B}$ drift terms which are much smaller than the $\bm{E}\times \bm{B}$ term. The equation of motion in this regime can be written in the following form:

\begin{equation}
    \begin{aligned}
        \frac{d \bm{r}}{d t} &= \frac{u_\parallel}{\gamma} + \bm{v}_E,\\
        \frac{d u_\parallel}{d t} &= \frac{q_s}{m_s} E_\parallel,\\
        \mu &\equiv \frac{m_s u_{\perp g}^2}{2 B}\sqrt{1- v_E^2 / c^2} = 0.
    \end{aligned}
\end{equation}
\noindent Here $\bm{v}_E$ is the three-velocity of the frame where $\bm{E}$ and $\bm{B}$ are parallel,

\begin{equation}
    \bm{v}_E = \frac{\bm{w}_E}{2 w_E^2 / c^2}\left(1 - \sqrt{1 - 4 w_E^2 / c^2}\right),
\end{equation}
\noindent with $\bm{w}_E / c = \bm{E}\times\bm{B} / (E^2 + B^2)$. Here we implicitly decomposed the velocity of particle into three different components:

\begin{equation}
    \bm{u} = u_\parallel \hat{\bm{B}} + \bm{v}_E \gamma + \bm{u}_{\perp g},
\end{equation}
\noindent where $u_{\perp g}$ is the particle velocity component responsible for its gyration. Notice, that we enforce the magnetic moment of the particle to be zero in the GCA regime. This assumption has physical justification: in pulsar magnetospheres bulk of the particles essentially reside at the zeroth Landau level, radiating their perpendicular momentum via synchrotron mechanism almost instantly; setting $\mu=0$ when particle switches from the conventional algorithm to GCA is similar to that process. GCA algorithm allows to significantly underresolve cold particle gyration close to the stellar surface as well as in the most of the magnetosphere, while still being able to model the strong synchrotron cooling for the high-energy particles treated with the conventional pusher.

In order to properly capture the regions where important kinetic plasma physics occurs (such as reconnection in the current layer), while still retaining the GCA approach for the bulk of the particles in the magnetosphere, we employ simple criteria for switching between the conventional Boris and GCA. If either of the following two conditions is satisfied, the full equation of motion of the particle is solved with the conventional algorithm:

\begin{equation}
    \begin{aligned}
        \gamma\beta \frac{m_s c^2}{|q_s| B} &> f_B \Delta x,\\
        \frac{E}{B} &> f_E.
    \end{aligned}
\end{equation}

\noindent Here $f_E$ and $f_B$ are dimensionless numbers which we typically choose to be $1$ and $0.95$ correspondingly. The first condition ensures that only particles with unresolved gyroradii are reduced to their corresponding guiding centers. The second condition ensures that the GCA is not applied in the regions of vanishing magnetic field where particle gyration is ill-defined.


\section{Dependence of the reconnection rate on the upstream motion}
\label{sec:psr-appendixC}

To test how the rate of magnetic reconnection depends on the bulk velocity of upstream plasma along the layer we perform 2D localized simulations. We start with a Harris equilibrium defined in the frame moving with a Lorentz factor $\Gamma$ along the current layer with respect to the lab frame. In the first set of simulations we keep $\sigma_{\rm up} / \Gamma = B^2 / (4\pi\rho_{\rm up} \Gamma) = 20$ constant ($\rho_{\rm up}$ is the upstream density in the lab frame) and vary $\Gamma$ (meaning that we also vary $\sigma_{\rm up}$). In the second set, we keep $\sigma_{\rm up} = 100$ \NEW{(as measured in the lab frame),} while, again, varying $\Gamma$. After about a light crossing time of the box, when the reconnection proceeds in the plasmoid-dominated stage (but before it shuts down because of the finite size of the box) we can measure the effective reconnection rate in two ways. We can either directly measure the upstream drift velocity towards the current layer, $\beta_{\rm in} = (\bm{E}\times\bm{B})_{\rm in}/B^2$, or we can compute how fast the magnetic energy is dissipated by evaluating the slope of the $d E_B / d t$ curve, where $E_B = \int_V d^3\bm{r} B^2/8\pi$ is the total magnetic energy contained in the box. Both of these methods unsurprisingly provide the same answer, so below we only report the direct measurement of $\beta_{\rm rec}$.

\begin{figure}[htb]
\centering
\includegraphics[width=0.6\columnwidth]{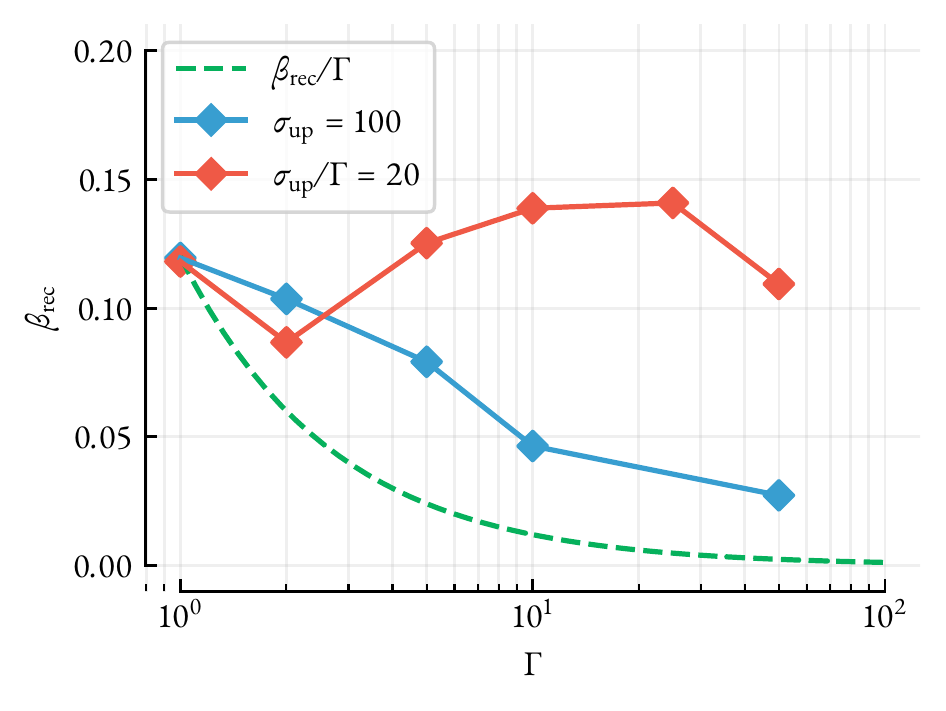}
\includegraphics[width=0.35\columnwidth]{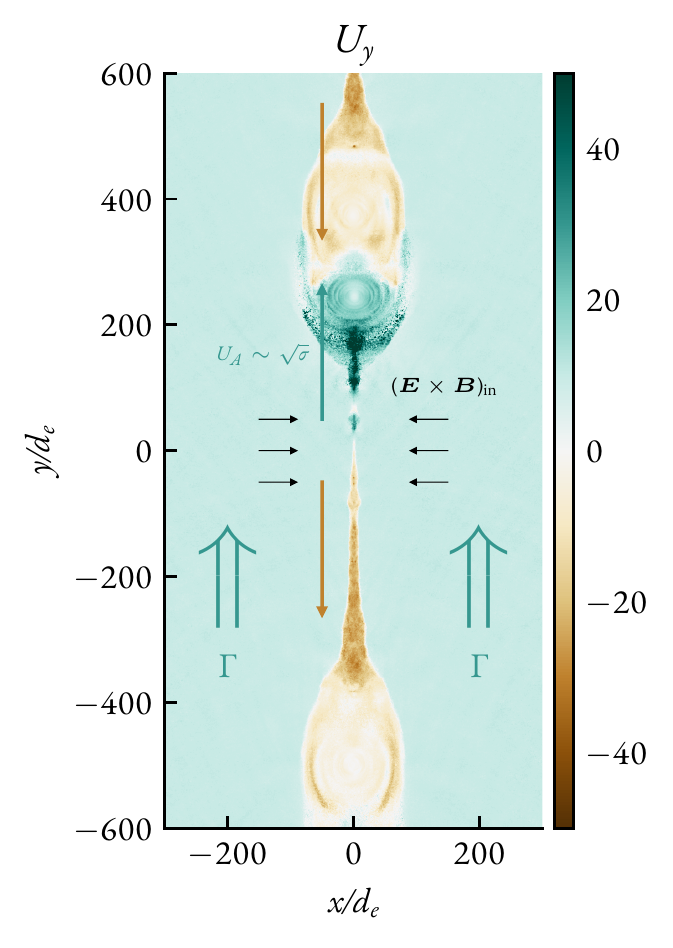}
\caption{{Left: reconnection rate, measured as $\beta_{\rm rec}=(\bm{E}\times\bm{B})_{\rm in}/B^2$ in the upstream for localized 2D simulations initialized with an upstream flying with a Lorentz factor of $\Gamma$ along the current layer. Green dashed line shows the naive expectation, that the rate, being universal in the proper frame of the upstream plasma, in the lab frame should scale as $1/\Gamma$. Right: zoom in to a region of our simulation with $\sigma_{\rm up}=100$, $\Gamma=10$. Plot shows the 4-velocity component of plasma along the sheet, $U_y$. Initially the upstream plasma is pushed in the $+y$ direction. In the current sheet, nevertheless, relativistic outflows move both directions, with the $x$-points in-between being static. }}
\label{fig:psr-appendixC-boost}
\end{figure}

In Figure~\ref{fig:psr-appendixC-boost} (left) we plot our measurements for the inflow velocity, $\beta_{\rm in}=\beta_{\rm rec}$ as a function of the bulk Lorentz factor of the upstream, $\Gamma$, for the two simulation series mentioned above. Red curve corresponds to the case when $\sigma_{\rm up}/\Gamma = 20$ is kept constant; in this series we see almost no variation in the inflow velocity, \NEW{the value of which lies} between $0.8$ and $0.14$. Notice, that keeping $\sigma_{\rm up}/\Gamma$ fixed means that we scale $\sigma_{\rm up}$ (in the lab frame) up with increasing $\Gamma$. In the case when $\sigma_{\rm up}$ is kept fixed we see a moderate decrease in $\beta_{\rm rec}$. 

\NEW{
To understand this effect, let us consider separately the velocities of relativistic outflows from the x-point (as shown in the right panel of Figure~\ref{fig:psr-appendixC-boost}). In the frame co-moving with the upstream plasma the velocities of these outflows should be symmetric w.r.t. the y-axis. Or in other words, $u'_\pm \approx u_\mathcal{A}'= \sqrt{\sigma'}$, where dashed quantities are measured in the frame co-moving with upstream plasma. Here $\sigma'\equiv \left(B'\right)^2/(4\pi\rho_e' c^2)$; ``$\pm$'' indicate the direction of motion along and opposite to the current layer (the y-axis). Boosting back to the lab frame we find: $u_+\approx u'_+\Gamma$, and $u_-\approx u'_-/\Gamma$, assuming the ``upstream'' frame moves along the y-axis with a Lorentz factor of $\Gamma\gg 1$. Cold upstream magnetization (not including bulk Lorentz factor) in the lab frame, $\sigma\equiv B^2/(4\pi \rho_e c^2)$, can be expressed as $\sigma=\sigma'/\Gamma$, since $B_{\rm up}=B_y$ does not transform, and $\rho_e=\Gamma \rho'_e$. Then the outflow velocities in the lab frame can be written as: $u_+\approx u_\mathcal{A} \Gamma^{3/2}$, and $u_-\approx u_\mathcal{A}\Gamma^{-1/2}$, where $u_\mathcal{A}=\sqrt{\sigma}$ is the relativistic Alfv\'en velocity in the lab frame. This can be also seen in the right panel of Figure~\ref{fig:psr-appendixC-boost}: outflow velocities along the y-axis measured in the lab frame are clearly asymmetric. When $\Gamma^{1/2}$ becomes comparable to $u_\mathcal{A}$, the outflow velocity in the direction opposite to the upstream boost, $u_-$, may become non-relativistic. In this case it is natural to expect the reconnection rate, defined in our case as $\beta_{\rm rec} \equiv v_{\rm in}/v_{\rm out} \approx v_{\rm in}/c$ to drop, since the effective outflow three-velocity, $v_{\rm out}$, can now be smaller than $c$.
} 


Arguments provided in this section are not exhaustive, and should rather be taken as possible explanations of empirical facts, observed in numerous simulations. Further investigation is definitely necessary to better understand the nature of the mechanism that controls the reconnection rate in these cases.

\section{Energy distribution of particles in the different parts of the simulation domain}
\label{sec:psr-appendixD}
For the simulation \texttt{R75\_ang0} magnetization parameter in the upstream, shown in Figure~\ref{fig:psr-sigma}, is close to $\sigma^{\rm LC}\sim 10^3$ (as shown with the line-out plot on the right) and drops to zero inside the current sheet, where the magnetic field vanishes. This means that the characteristic energies to which particles can get accelerated in the current sheet are comparable to $\sigma m_e c^2$ (as will be demonstrated shortly).

\begin{figure*}[htb]
\centering
\includegraphics[width=0.75\columnwidth,trim={10 0 10 5},clip]{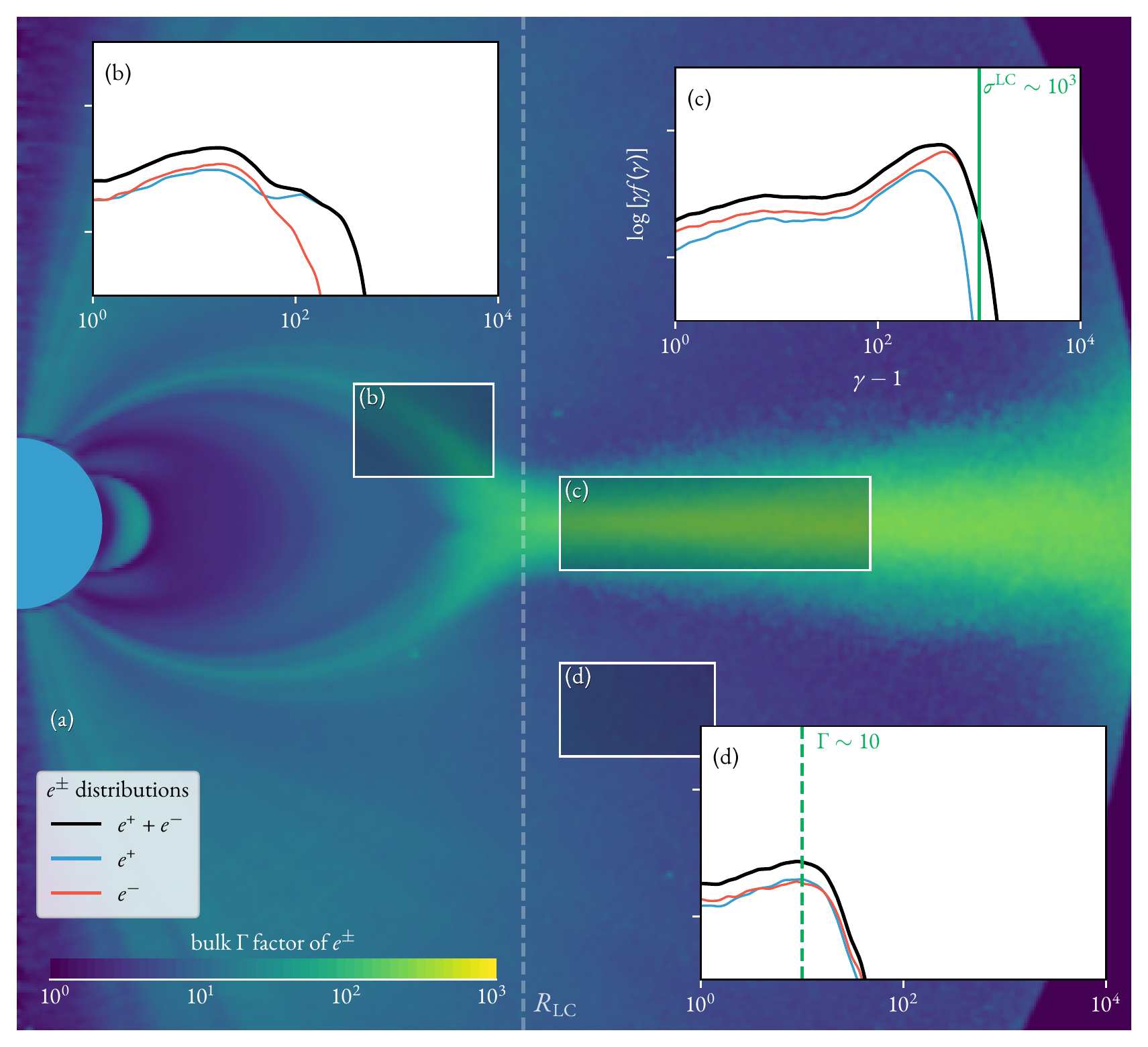}
\caption{Distribution functions of electrons and positrons measured in different locations of our simulation shown on an azimuthally averaged poloidal slice. {Synchrotron cooling in this particular simulation (\texttt{R75\_ang0\_gr2}) is weak.} Bulk Lorentz factor is shown in panel (a). Panels (b), (c), and (d) show distribution functions in the separatrix, the current layer, and the upstream, respectively. Plasma in the upstream is typically cold with a bulk outflow Lorentz factor of a few. In the current sheet the bulk motion of plasma is dictated by the dynamics of reconnection, and the characteristic outflow velocities are comparable to the Alfv\'en speed with a Lorentz factor $\sim\sqrt{\sigma}$. As expected, the highest energy particles are produced in the current layer where magnetic reconnection occurs.}
\label{fig:psr-spatial_spectra}
\end{figure*}

Let us consider distribution functions for $e^\pm$ in different regions of our simulation \texttt{R75\_ang0}, as shown in Figure~\ref{fig:psr-spatial_spectra}. On a poloidal slice in panel \ref{fig:psr-spatial_spectra}a we show the bulk Lorentz factor of the plasma, $\Gamma = \sqrt{1 + \bm{U}^2/c^2}$, which is computed using the bulk four-velocity, $\bm{U} = \int \bm{u}f(\bm{u})d^3\bm{r}$. Panels \ref{fig:psr-spatial_spectra}b, \ref{fig:psr-spatial_spectra}c, and \ref{fig:psr-spatial_spectra}d show distribution functions of the electrons and positrons in three different regions: in the separatrix (last closed field line), in the current sheet, and in the upstream correspondingly. \emph{Upstream} plasma is relatively cold; {electrons and positron flow outwards along the magnetic field lines, gaining characteristic bulk Lorentz factors of $\Gamma \sim 10$ (Figure~\ref{fig:psr-spatial_spectra}d).} On the \emph{separatrix} (Figure~\ref{fig:psr-spatial_spectra}b) both electrons and positrons from the surface are marginally accelerated by an unscreened electric field, gaining energies close to $\langle\gamma \rangle\sim 10\text{-}100$. This region also hosts hot electrons returning from the Y-point, which is evident from the excess of electrons at Lorentz factors of a few $10^2$ shown in Figure~\ref{fig:psr-spatial_spectra}b.

These two regions, the upstream and the separatrix, act as an ``intake'' and an ``exhaust'' for the \emph{current sheet}. It is important to note that particles in these regions are generally exposed to a substantial orthogonal magnetic field component. This means that high-energy particles from the current sheet cannot exist there for timescales longer than the short cooling time. Thus, the main role in shaping the observed $\gamma$-ray emission is played by the current sheet, where constant release of magnetic energy sustainably accelerates particles in x-lines, where the magnetic field strength is zero (the cooling is non-existent).

In the current sheet (Figure~\ref{fig:psr-spatial_spectra}c) we see a substantial non-thermal particle population, extending to $\gamma\sim \sigma\sim 500\text{-}1000$. The bulk motion of the current sheet, on the other hand, has a Lorentz factor of $\Gamma\sim 10\text{-}100$, consistent with characteristic velocities of relativistic flows along the current sheet during the reconnection, $\Gamma\sim \sqrt{\sigma}$ (which corresponds to the Alfv\'en speed).\footnote{Slight enhancement is due to the fact that there is a bulk $\bm{E}\times\bm{B}$ outflow of the current sheet, similar to the wind in the upstream.} Notice also, that the distributions of electrons and positrons are slightly different: positrons extend to slightly higher energies. This is due to the fact that the electric field generated during reconnection ``pushes'' the electrons opposite to the global $\bm{E}\times\bm{B}$ drift, which is charge-invariant and directed radially outward (in Figure~\ref{fig:psr-pulsar3d}a the reconnection electric field is pointing in the out-of-plane direction {and has both toroidal and radial components}). This difference in the electrons and positrons has been observed in earlier simulations \citep[see, e.g.,][]{2015MNRAS.448..606C}; however, it almost vanishes for large obliquity angles, $\chi$, which we also see in our simulations.

\bibliography{references}{}

\begin{thebibliography}{}
\expandafter\ifx\csname natexlab\endcsname\relax\def\natexlab#1{#1}\fi
\providecommand{\url}[1]{\href{#1}{#1}}
\providecommand{\dodoi}[1]{doi:~\href{http://doi.org/#1}{\nolinkurl{#1}}}
\providecommand{\doeprint}[1]{\href{http://ascl.net/#1}{\nolinkurl{http://ascl.net/#1}}}
\providecommand{\doarXiv}[1]{\href{https://arxiv.org/abs/#1}{\nolinkurl{https://arxiv.org/abs/#1}}}

\bibitem[{{Abdo} {et~al.}(2013){Abdo}, {Ajello}, {Allafort}, {Baldini},
  {Ballet}, {Barbiellini}, {Baring}, {Bastieri}, {Belfiore}, {Bellazzini}, \&
  et~al.}]{2013ApJS..208...17A}
{Abdo}, A.~A., {Ajello}, M., {Allafort}, A., {et~al.} 2013, \apjs, 208, 17,
  \dodoi{10.1088/0067-0049/208/2/17}

\bibitem[{{Akhiezer} \& {Berestetskij}(1985)}]{1985quel.book.....A}
{Akhiezer}, A., \& {Berestetskij}, V.~B. 1985, {Quantum Electrodynamics}

\bibitem[{{Bacchini} {et~al.}(2020){Bacchini}, {Ripperda}, {Philippov}, \&
  {Parfrey}}]{2020ApJS..251...10B}
{Bacchini}, F., {Ripperda}, B., {Philippov}, A.~A., \& {Parfrey}, K. 2020,
  \apjs, 251, 10, \dodoi{10.3847/1538-4365/abb604}

\bibitem[{{Belyaev}(2015)}]{2015MNRAS.449.2759B}
{Belyaev}, M.~A. 2015, \mnras, 449, 2759, \dodoi{10.1093/mnras/stv468}

\bibitem[{{Cerutti} {et~al.}(2015){Cerutti}, {Philippov}, {Parfrey}, \&
  {Spitkovsky}}]{2015MNRAS.448..606C}
{Cerutti}, B., {Philippov}, A., {Parfrey}, K., \& {Spitkovsky}, A. 2015,
  \mnras, 448, 606, \dodoi{10.1093/mnras/stv042}

\bibitem[{{Cerutti} \& {Philippov}(2017)}]{2017A&A...607A.134C}
{Cerutti}, B., \& {Philippov}, A.~A. 2017, \aap, 607, A134,
  \dodoi{10.1051/0004-6361/201731680}

\bibitem[{{Cerutti} {et~al.}(2020){Cerutti}, {Philippov}, \&
  {Dubus}}]{2020A&A...642A.204C}
{Cerutti}, B., {Philippov}, A.~A., \& {Dubus}, G. 2020, \aap, 642, A204,
  \dodoi{10.1051/0004-6361/202038618}

\bibitem[{{Cerutti} {et~al.}(2016){Cerutti}, {Philippov}, \&
  {Spitkovsky}}]{2016MNRAS.457.2401C}
{Cerutti}, B., {Philippov}, A.~A., \& {Spitkovsky}, A. 2016, \mnras, 457, 2401,
  \dodoi{10.1093/mnras/stw124}

\bibitem[{{Cerutti} {et~al.}(2014){Cerutti}, {Werner}, {Uzdensky}, \&
  {Begelman}}]{2014ApJ...782..104C}
{Cerutti}, B., {Werner}, G.~R., {Uzdensky}, D.~A., \& {Begelman}, M.~C. 2014,
  \apj, 782, 104, \dodoi{10.1088/0004-637X/782/2/104}

\bibitem[{{Chen} \& {Beloborodov}(2014)}]{2014ApJ...795L..22C}
{Chen}, A.~Y., \& {Beloborodov}, A.~M. 2014, \apjl, 795, L22,
  \dodoi{10.1088/2041-8205/795/1/L22}

\bibitem[{{Contopoulos} {et~al.}(1999){Contopoulos}, {Kazanas}, \&
  {Fendt}}]{1999ApJ...511..351C}
{Contopoulos}, I., {Kazanas}, D., \& {Fendt}, C. 1999, \apj, 511, 351,
  \dodoi{10.1086/306652}

\bibitem[{{Deutsch}(1955)}]{1955AnAp...18....1D}
{Deutsch}, A.~J. 1955, Annales d'Astrophysique, 18, 1

\bibitem[{{Goldreich} \& {Julian}(1969)}]{1969ApJ...157..869G}
{Goldreich}, P., \& {Julian}, W.~H. 1969, \apj, 157, 869,
  \dodoi{10.1086/150119}

\bibitem[{{Gruzinov}(2005)}]{2005PhRvL..94b1101G}
{Gruzinov}, A. 2005, \prl, 94, 021101, \dodoi{10.1103/PhysRevLett.94.021101}

\bibitem[{{Guo} {et~al.}(2014){Guo}, {Li}, {Daughton}, \&
  {Liu}}]{2014PhRvL.113o5005G}
{Guo}, F., {Li}, H., {Daughton}, W., \& {Liu}, Y.-H. 2014, \prl, 113, 155005,
  \dodoi{10.1103/PhysRevLett.113.155005}

\bibitem[{{Guo} {et~al.}(2021){Guo}, {Li}, {Daughton}, {Li}, {Kilian}, {Liu},
  {Zhang}, \& {Zhang}}]{2021ApJ...919..111G}
{Guo}, F., {Li}, X., {Daughton}, W., {et~al.} 2021, \apj, 919, 111,
  \dodoi{10.3847/1538-4357/ac0918}

\bibitem[{{Hakobyan} {et~al.}(2021){Hakobyan}, {Petropoulou}, {Spitkovsky}, \&
  {Sironi}}]{2021ApJ...912...48H}
{Hakobyan}, H., {Petropoulou}, M., {Spitkovsky}, A., \& {Sironi}, L. 2021,
  \apj, 912, 48, \dodoi{10.3847/1538-4357/abedac}

\bibitem[{{Hakobyan} {et~al.}(2019){Hakobyan}, {Philippov}, \&
  {Spitkovsky}}]{2019ApJ...877...53H}
{Hakobyan}, H., {Philippov}, A., \& {Spitkovsky}, A. 2019, \apj, 877, 53,
  \dodoi{10.3847/1538-4357/ab191b}

\bibitem[{Hakobyan \& Spitkovsky(2020)}]{tristanv2}
Hakobyan, H., \& Spitkovsky, A. 2020, Tristan-MP v2, multi-species
  particle-in-cell plasma code:
  \texttt{github.com/PrincetonUniversity/tristan-mp-v2}.
\newblock \url{https://princetonuniversity.github.io/tristan-v2/}

\bibitem[{{Hu} \& {Beloborodov}(2021)}]{2021arXiv210903935H}
{Hu}, R., \& {Beloborodov}, A.~M. 2021, arXiv e-prints, arXiv:2109.03935.
\newblock \doarXiv{2109.03935}

\bibitem[{{Kagan} {et~al.}(2016){Kagan}, {Nakar}, \&
  {Piran}}]{2016ApJ...826..221K}
{Kagan}, D., {Nakar}, E., \& {Piran}, T. 2016, \apj, 826, 221,
  \dodoi{10.3847/0004-637X/826/2/221}

\bibitem[{{Kalapotharakos} {et~al.}(2018){Kalapotharakos}, {Brambilla},
  {Timokhin}, {Harding}, \& {Kazanas}}]{2018ApJ...857...44K}
{Kalapotharakos}, C., {Brambilla}, G., {Timokhin}, A., {Harding}, A.~K., \&
  {Kazanas}, D. 2018, \apj, 857, 44, \dodoi{10.3847/1538-4357/aab550}

\bibitem[{{Lyubarskii}(1996)}]{1996A&A...311..172L}
{Lyubarskii}, Y.~E. 1996, \aap, 311, 172

\bibitem[{{Marelli} {et~al.}(2011){Marelli}, {De Luca}, \&
  {Caraveo}}]{2011ApJ...733...82M}
{Marelli}, M., {De Luca}, A., \& {Caraveo}, P.~A. 2011, \apj, 733, 82,
  \dodoi{10.1088/0004-637X/733/2/82}

\bibitem[{{Michel}(1973)}]{1973ApJ...180L.133M}
{Michel}, F.~C. 1973, \apjl, 180, L133, \dodoi{10.1086/181169}

\bibitem[{{Petropoulou} \& {Sironi}(2018)}]{2018MNRAS.481.5687P}
{Petropoulou}, M., \& {Sironi}, L. 2018, \mnras, 481, 5687,
  \dodoi{10.1093/mnras/sty2702}

\bibitem[{{Philippov} \& {Spitkovsky}(2014)}]{2014ApJ...785L..33P}
{Philippov}, A.~A., \& {Spitkovsky}, A. 2014, \apj, 785, L33,
  \dodoi{10.1088/2041-8205/785/2/L33}

\bibitem[{{Philippov} \& {Spitkovsky}(2018)}]{PSAS18}
---. 2018, \apj, 855, 94, \dodoi{10.3847/1538-4357/aaabbc}

\bibitem[{{Philippov} {et~al.}(2015){Philippov}, {Spitkovsky}, \&
  {Cerutti}}]{2015ApJ...801L..19P}
{Philippov}, A.~A., {Spitkovsky}, A., \& {Cerutti}, B. 2015, \apj, 801, L19,
  \dodoi{10.1088/2041-8205/801/1/L19}

\bibitem[{{Ripperda} {et~al.}(2022){Ripperda}, {Liska}, {Chatterjee}, {Musoke},
  {Philippov}, {Markoff}, {Tchekhovskoy}, \& {Younsi}}]{Ripperda2022}
{Ripperda}, B., {Liska}, M., {Chatterjee}, K., {et~al.} 2022, \apjl, 924, L32

\bibitem[{{Ruderman} \& {Sutherland}(1975)}]{1975ApJ...196...51R}
{Ruderman}, M.~A., \& {Sutherland}, P.~G. 1975, \apj, 196, 51,
  \dodoi{10.1086/153393}

\bibitem[{{Sironi}(2022)}]{2022PhRvL.128n5102S}
{Sironi}, L. 2022, \prl, 128, 145102, \dodoi{10.1103/PhysRevLett.128.145102}

\bibitem[{{Sironi} \& {Spitkovsky}(2014)}]{2014ApJ...783L..21S}
{Sironi}, L., \& {Spitkovsky}, A. 2014, \apjl, 783, L21,
  \dodoi{10.1088/2041-8205/783/1/L21}

\bibitem[{{Sturrock}(1971)}]{1971ApJ...164..529S}
{Sturrock}, P.~A. 1971, \apj, 164, 529, \dodoi{10.1086/150865}

\bibitem[{{Tchekhovskoy} {et~al.}(2013){Tchekhovskoy}, {Spitkovsky}, \&
  {Li}}]{2013MNRAS.435L...1T}
{Tchekhovskoy}, A., {Spitkovsky}, A., \& {Li}, J.~G. 2013, \mnras, 435, L1,
  \dodoi{10.1093/mnrasl/slt076}

\bibitem[{{Timokhin}(2006)}]{2006MNRAS.368.1055T}
{Timokhin}, A.~N. 2006, \mnras, 368, 1055,
  \dodoi{10.1111/j.1365-2966.2006.10192.x}

\bibitem[{{Timokhin}(2010)}]{2010MNRAS.408.2092T}
---. 2010, \mnras, 408, 2092, \dodoi{10.1111/j.1365-2966.2010.17286.x}

\bibitem[{{Timokhin} \& {Harding}(2015)}]{2015ApJ...810..144T}
{Timokhin}, A.~N., \& {Harding}, A.~K. 2015, \apj, 810, 144,
  \dodoi{10.1088/0004-637X/810/2/144}

\bibitem[{{Timokhin} \& {Harding}(2019)}]{2019ApJ...871...12T}
---. 2019, \apj, 871, 12, \dodoi{10.3847/1538-4357/aaf050}

\bibitem[{{Uzdensky} \& {Spitkovsky}(2014)}]{2014ApJ...780....3U}
{Uzdensky}, D.~A., \& {Spitkovsky}, A. 2014, \apj, 780, 3,
  \dodoi{10.1088/0004-637X/780/1/3}

\bibitem[{{Werner} \& {Uzdensky}(2021)}]{2021JPlPh..87f9013W}
{Werner}, G.~R., \& {Uzdensky}, D.~A. 2021, Journal of Plasma Physics, 87,
  905870613, \dodoi{10.1017/S0022377821001185}

\bibitem[{{Werner} {et~al.}(2018){Werner}, {Uzdensky}, {Begelman}, {Cerutti},
  \& {Nalewajko}}]{2018MNRAS.473.4840W}
{Werner}, G.~R., {Uzdensky}, D.~A., {Begelman}, M.~C., {Cerutti}, B., \&
  {Nalewajko}, K. 2018, \mnras, 473, 4840, \dodoi{10.1093/mnras/stx2530}

\bibitem[{{Werner} {et~al.}(2016){Werner}, {Uzdensky}, {Cerutti}, {Nalewajko},
  \& {Begelman}}]{2016ApJ...816L...8W}
{Werner}, G.~R., {Uzdensky}, D.~A., {Cerutti}, B., {Nalewajko}, K., \&
  {Begelman}, M.~C. 2016, \apjl, 816, L8, \dodoi{10.3847/2041-8205/816/1/L8}

\bibitem[{{Zhang} {et~al.}(2021){Zhang}, {Sironi}, \&
  {Giannios}}]{2021ApJ...922..261Z}
{Zhang}, H., {Sironi}, L., \& {Giannios}, D. 2021, \apj, 922, 261,
  \dodoi{10.3847/1538-4357/ac2e08}

\end{thebibliography}
\bibliographystyle{aasjournal}

\end{document}